\documentclass[a4paper,fleqn,usenatbib,twocolumn]{mnras}
\usepackage{newtxtext,newtxmath}
\usepackage[T1]{fontenc}
\usepackage{ae,aecompl}
\usepackage{graphicx}
\usepackage{amsmath}	
\usepackage{subfig}
\usepackage{multirow}
\usepackage{tabularx}
\usepackage{array}
\usepackage{booktabs} 
\newcolumntype{P}[1]{>{\centering\arraybackslash}p{#1}}
\newcommand{\be}{\begin{equation}}
\newcommand{\ee}{\end{equation}}
\usepackage{calligra} 
\DeclareMathAlphabet{\mathcalligra}{T1}{calligra}{m}{n} 
\DeclareFontShape{T1}{calligra}{m}{n}{<->s*[2.2]callig15}{}   

\newcommand{\mach}{\mathcal{M}}

\title[Inspirals in gas discs]{Evolution of gas disc{\textendash}embedded intermediate mass ratio inspirals in the LISA band}
\author[A. Derdzinski]{A.~Derdzinski$^{1,2}\thanks{E-mail: andrea@ics.uzh.ch}$, D. D'Orazio$^{2}$, P. Duffell$^{3}$,  Z. Haiman$^{1}$, A. MacFadyen$^{4}$\\
$^{1}$Department of Astronomy, Columbia University, New York, NY, 10027, USA\\
$^{2}$Center for Theoretical Astrophysics and Cosmology, Institute for Computational Science, University of Zurich, \\
Winterthurerstrasse 190, 8057 Zurich, Switzerland\\
$^{3}$Department of Astronomy, Harvard University, 60 Garden Street Cambridge, MA 01238, USA\\
$^{4}$Center for Cosmology and Particle Physics, Physics Department, New York University, New York, NY 10003,USA}

\begin{document}
\date{Received / Accepted}
\pagerange{\pageref{firstpage}--\pageref{lastpage}} \pubyear{2017}

\maketitle
\label{firstpage}

\begin{abstract}
Among the potential milliHz gravitational wave (GW) sources for the upcoming space-based interferometer LISA are extreme- or intermediate-mass ratio inspirals (EMRI/IMRIs). These events involve the coalescence of supermassive black holes in the mass range $10^5 M_{\odot} \lesssim M \lesssim 10^7 M_{\odot}$ with companion BHs of much lower masses. A subset of E/IMRIs are expected to occur in the accretion discs of active galactic nuclei (AGN), where torques exerted by the disc can interfere with the inspiral and cause a phase shift in the GW waveform. Here we use a suite of two-dimensional hydrodynamical simulations with the moving-mesh code DISCO to present a systematic study of disc torques. We measure torques on an inspiraling BH and compute the corresponding waveform deviations as a function of the binary mass ratio $q\equiv M_2/M_1$, the disc viscosity ($\alpha$), and gas temperature (or equivalently Mach number; $\mach$). We find that the absolute value of the gas torques is within an order of magnitude of previously determined planetary migration torques, but their precise value and sign depends non-trivially on the combination of these parameters. The gas imprint is detectable by LISA for binaries embedded in AGN discs with surface densities above $\Sigma_0\ge10^{4-6} \rm \, g cm^{-2}$, depending on $q$, $\alpha$ and $\mach$.  Deviations are most pronounced in discs with higher viscosities, and for E/IMRIs detected at frequencies where LISA is most sensitive. Torques in colder discs exhibit a noticeable dependence on the GW-driven inspiral rate as well as strong fluctuations at late stages of the inspiral. Our results further suggest that LISA may be able to place constraints on AGN disc parameters and the physics of disc-satellite interaction. 
\end{abstract}


\section{Introduction} 
\label{sec:introduction}

 In the 2030s we expect to detect binary mergers involving massive black holes (MBHs;
in the mass range $M_{\rm BH}\!\sim\! 10^4\!-\!10^7 M_{\odot}$) across the universe  with the space-based gravitational wave (GW) detector LISA \citep{Seoane2017}.
LISA will also detect extreme and intermediate mass ratio mergers 
(termed EMRIs, $q\equiv M_2/M_1\!\lesssim\!10^{-4}$; or IMRIs, $q\!\approx\!10^{-3}\!-\!10^{-4}$) up to a redshift $z\!\sim\!4$.

Many near-equal mass MBH mergers in the LISA band may occur in gaseous environments, 
given that the galactic mergers that lead to the eventual formation of a MBH 
binary often carry a fresh supply of gas into the post-merger galactic nucleus (see \citealt{mayer2013} for a review). 
Likewise, "gas-embedded" E/IMRIs may frequently occur in the accretion discs expected in active galactic nuclei (AGN) 
in which an MBH is surrounded by a thin, dense accretion disc. 
Compact objects in the nucleus can eventually align their orbits with the disc, provided a sufficient number of disc-crossing intersections (e.g. \citealt{Syer1991,1993ApJ...409..592A,1995MNRAS.275..628R,2012ApJ...758...51J,McKernan2012,2016MNRAS.460..240K,1993ApJ...409..592A,Fabj2020,2018MNRAS.476.4224P,MacleodLin2020}), or such events may arise from in-situ star formation in the disc that leaves compact remnants
\citep{GoodmanTan2004,Levin2007}.
Embedded stars and BHs can subsequently accrete, migrate, and merge with each other  \citep{Bellovary2016,Tagawa+2020,2019ApJ...878...85S,2020arXiv200411936S},
before eventually merging with the central MBH. 
While the event rate for E/IMRIs has previously been estimated in dry nuclei (e.g. \citealt{BarausseCardosoPani2015}), 
the rate may be considerably higher when including formation pathways in accretion discs. 
The recent discovery of stellar-mass BH mergers by the ground based interferometer LIGO 
has stimulated work on such mergers that may be occurring in AGN discs, showing that they might contribute to LIGO events \citep{McKernan2014,Bartos2017,McKernan2017,2020A&A...638A.119G,Stone2017}.  Subsequent work has suggested that some of the LIGO events, based on their high chirp masses, may indeed have formed via this channel - large masses are expected both because of the preferential capture of heavier BHs by the disc~\citep{Yang+2019a} and because repeated mergers are common and lead to hierarchical build-up~\citep{Yang+2019b}.  The high effective spin provides additional support for this channel~\citep{Gayathri+2020}   
 as well as events with more unequal mass ratios (e.g. GW190412, see \citealt{Tagawa2020b}) which have been predicted to occur naturally via dynamical interactions in a dense, gaseous environment \citep{2020MNRAS.494.1203M,Tagawa+2020,2020arXiv200411936S}. 
These results suggest that BH mergers indeed occur in gas discs, and that they could constitute the building blocks of IMBHs~\citep{McKernan2012,Tagawa+2020}. 
Recent semi-analytical estimates of in-situ star formation suggest that the accretion of 
embedded compact objects may also contribute substantially to the growth of MBHs \citep{Dittmann2019}.

In general, whenever gas is present around a coalescing MBH binary, the torques exerted by the gas can influence the inspiral rate.  For a near-equal mass binary 
in the late inspiral stage, 
these torques are noticeable 
only at extreme gas densities, comparable to those expected if the binary is embedded in a common envelope phase~\citep{Antoni+2019,ChenShen2019}.
For extreme- and intermediate- mass ratio systems, the GW torques are 
weaker and gas torques on the lower-mass companion are stronger.
As a result, for a long-lived E/IMRI detected and monitored by several years with the low-frequency GW detector LISA, gas torques can become more comparable to (albeit still well below) GW torques,
and impart a detectable imprint on the GW waveform \citep{Yunes2011,Kocsis2011,Derdzinski2019}. 
This presents a novel and unique opportunity for LISA to detect environmental influence in a 
MBH waveform and probe accretion disc physics.

A longstanding hurdle in 
numerically evaluating this effect is that gas torques are 
subtle, remain poorly understood, and show a complex dependence on system parameters (see \S~\ref{sec:prev} below).  
The response of a disc to an embedded satellite can be highly nonlinear
\citep{BaruteauMasset2013}, making simulations necessary to quantify it, 
and the resulting torque can be remarkably sensitive to small changes in the satellite mass or disc parameters \citep{Duffell2015}.
Solving for the disc torques requires especially careful considerations of numerical resolution, boundary effects and transients (i.e. achieving steady-state behaviour),  as well as the accretion of both mass and momentum by the migrating BH~\citep{Tang2017, Munoz2018, Moody+2019, Munoz+2020}.  
For a large range of mass ratios, simulations show that disc torques may either help or hinder the binary merger \citep{Duffell2020}.

In a previous paper (\citealt{Derdzinski2019}; hereafter Paper I), we simulated the gas response to 
an embedded IMRI (with mass ratio $q=10^{-3}$) in order to measure its detectability in the GW signal. We found that the disc torques slow 
down, rather than speed up the inspiral, due to a critical contribution to the torque that comes 
from the asymmetry in the gas morphology near the migrator (near or within the BH's Hill sphere). We estimated that the resulting deviation in the GW waveform, 
which comprises of a slow drift in the accumulated GW phase, is detectable if the IMRI resides in a disc with 
surface densities above $\Sigma_0\gtrsim10^{3-4} \rm g \, cm^{-2}$.

In Paper I, we considered only a single set of binary system and disc parameters.
In order to extract meaningful information from environmental imprints on a GW waveform, we must 
understand the range of possible effects. In particular, we need to know how the torques and the 
corresponding waveform deviations depend on system parameters (mass ratio, eccentricity, inclination, spins) and the source environment (in our case, properties of the disc such as density, temperature or viscosity). 
In the present paper, we further explore the scaling of the torque with a subset of these parameters in the regime where GW-emission is dominant. 

This work is motivated by the idea that, provided we understand how gas torques impact a GW inspiral, 
GWs can be used as a tool for providing measurements of AGN disc properties, and to improve our understanding of the physics of migration. 
To this end, we study disc torques as a function of the GW inspiral rate in order to isolate the effect of 
the inspiral, and to assess how the disc torques may evolve differently for different combinations of parameters.  We restrict this study to include only three parameters - namely the binary mass ratio  ($q$) and two of the disc parameters: temperature (or equivalently Mach number; $\mach$) and viscosity (parameterised by the Shakura-Sunyaev parameter $\alpha$).  
As in Paper I, we measure the torques directly in hydrodynamical simulations over a limited range of binary separations, and then extrapolate these measurements to cover the final coalescence of a physical IMRI, covering several years of LISA observations.
Scaling our simulations to physical parameters,
we calculate the detectability of the gas imprint on the GW signal across our set of simulated mass ratios, 
and place constraints on the minimum AGN disc density required to detect the gas-induced deviations on the GW waveform. We show that 
the detectability of these deviations depends on the mass ratio and also the stage at which the inspiral is observed;  we also find that the time-evolution of the gas torques in the LISA window depends non-trivially on $q$, $\mach$, and $\alpha$.

This paper is organised as follows.  
In \S\ref{sec:prev}, we begin by summarising prior work. 
In \S\ref{sec:methods}, we describe our numerical approach, including the hydrodynamical simulations and the range of simulated parameters.
In \S\ref{sec:results}, we present our results, measuring and analysing the gas torques and their dependence on each parameter.
In \S\ref{sec:LISA}, we apply our results to LISA binaries, calculating the detectability of the gas-induced deviations in the waveform. 
Finally, we discuss our results in \S\ref{sec:discussion}, and summarize our conclusions in \S\ref{sec:conclusions}.

\section{Previous Work}
\label{sec:prev}

Here we summarize (i) applicable work on planetary torques for intermediate mass ratio systems, 
(ii) analytical work on disc torques on GW inspirals, and (iii) our previous paper that combines the two with simulations.

To understand the evolution of a low-mass satellite embedded in a gas disc, the large majority of the work to date has been done in the context of planetary migration in protoplanetary discs. The embedded 
satellite perturbs the disc non-axisymmetrically, and these perturbations back-react on 
the satellite's orbit. The perturbations include spiral density waves and gaps where streams of gas
can flow on horseshoe orbits around the satellite (see e.g. \citealt{BaruteauMasset2013} for a comprehensive review).
Historically, two distinct regimes have been identified and known as Type~I and Type~II
migration \citep{Ward97}, determined by the mass of the embedded satellite, as well as by the disc temperature and 
viscosity.
In the Type~I regime (for mass ratios $q\lesssim10^{-4}$, for typical disc temperatures and viscosities), the disc response is linear, and the migration rate 
can be predicted analytically
\citep{GT80} and described
with simple formulae (such as in \citealt{Tanaka2002}) that have been confirmed and calibrated with two- and three-dimensional simulations (e.g., \citealt{Dong2011, DAngelo2010, Duffell2012})
although these predictions assume that the disc is 
locally isothermal \citep{Paard2006}.
In the Type~II regime (or gap-opening regime) 
(typically $q\gtrsim10^{-4}$),
the disc response becomes nonlinear and the secondary begins to carve a low-density annular gap. 
The migration rate typically scales with some fraction of the local viscous rate (e.g. \citealt{Edgar2007}) 
and is proportional to the gas density in the gap \citep{Kanagawa2018}. However, the migration rate depends on disc parameters, and can even switch sign for certain combinations \citep{Duffell2014,Duffell2015}. 
One distinct regime is when the local disc mass is much smaller than the satellite's mass. Disc torques in this regime are lowered significantly, causing migration to slow down~\citep{SyerClarke1995,Duffell2014}.  Note that the E/IMRIs considered in our study are in this regime.
In summary, in both the Type~I and Type~II regimes,  the disc response, especially in the co-orbital region 
of the satellite, and the resulting torques on the satellite, are sensitive to disc parameters and to the equation of state. 
This makes predictions for real systems (whose parameters are unknown) difficult to make~\citep{Kley2012}.

Despite these caveats, analytical predictions are convenient and are often utilised in the literature. In the context of BHs embedded in gas discs, prior work has estimated the impact of gas torques on LISA sources in order to assess their 
detectability in the GW waveform \citep{Yunes2011,Kocsis2011,Barausse+2014}. These 
analytic studies adopted the gas torques as in the planet literature, and added these linearly to the effective GW torques.  They  have concluded that  gas torques on EMRIs, which can spend up to $O(10^5)$ cycles in the LISA band, are 
potentially detectable if embedded in thin, near-Eddington accretion discs
(i.e. discs whose steady-state accretion rate corresponds to the Eddington luminosity of the central SMBH, at a radiative efficiency of $\sim10\%$).

In Paper I, we improved on these analytic estimates by
performing the first direct measurement of gas disc torques on a gas-embedded IMRI, using two-dimensional hydrodynamical simulations. Unlike in planetary migration studies, the satellite was assumed to follow a GW-driven inspiral, and, unlike in the analytic estimates, we did not assume that that GW and disc torques can be added linearly.
Contrary to analytical estimates,
we found that a disc-embedded IMRI experiences \emph{outward} gas torques that hinder
the inspiral. This was shown for a single set of system and disc parameters: a binary mass ratio $q=10^{-3}$, viscosity parameter $\alpha=0.03$ and Mach number $\mach = 20$ 
(see below for the definitions of $\alpha$ and $\mach$).

Paper~I was the first to address how these torques may evolve during a GW-driven inspiral and to 
calculate their detectability directly from simulations.
This demonstrated a proof--of--concept for an optimal case: a $10^{-3}$ mass-ratio inspiral is 
relatively loud
(compared to EMRIs), and chirps substantially as it approaches merger.  These two 
qualities are paramount 
to detect the corresponding waveform deviations,
and to use the frequency-dependence of these deviations to distinguish them from variations in system parameters 
in order to securely identify them as environmental effects.

In the present work, we follow up on Paper~I and explore how the detectability of disc torques in GW waveforms depends on system or disc parameters.   Changing the mass ratio of the system will affect the gas dynamics (and the torque experienced by the secondary) as well as the GW evolution of the binary in the LISA band (and resulting detectability of the gas imprint).
We expand on the IMRI parameter space over an order of magnitude of mass ratio, demonstrating that reducing the secondary mass leads to different outcomes in torque evolution
and correspondingly more stringent constraints on detectability. 
We also then explore how these results depend on disc viscosity and Mach number. 
With a range of simulations we show that whether or not torques accelerate or hinder the GW-driven inspiral, and whether the resulting gas imprint is detectable in the GW data stream, will depend on the combinations of these parameters.

\section{Simulation Setup}
\label{sec:methods}

We use the moving-mesh hydrodynamics grid code {\sc DISCO} \citep{Duffell2016} to model a thin two-dimensional, viscous disc around an MBH with a low-mass satellite BH embedded in the disc, following a GW-driven inward migration ("inspiral").
In this section, we describe our scale-free numerical 
approach for the disc, the prescribed orbit of the migrator, as well as how we measure the torques.

\subsection{Disc model}
\label{sec:discmodel}

The simulation setup is the same as in Paper I, with slight modifications to the domain size and spatial
resolution. 
We model a vertically-integrated, near-Keplerian disc,
parameterised by a constant aspect ratio $h/r \equiv \mach^{-1}$, where $h$ is the disc scale 
height, $r$ is the distance to the central MBH 
held at the origin,
and $\mach$ is the Mach number for the azimuthal velocity
 $v_{\phi} = \sqrt{GM_1/r}$ around the central MBH of mass $M_1$.  For thin and cold discs, the latter is equivalent to specifying the local sound speed; 
$c_s = v_{\phi}/\mach$.  We adopt a constant $\alpha$-law 
prescription for the viscosity, such that the kinematic viscosity is set by $\nu (r) = \alpha c_s(r) h(r)$. 
We force the disc to be locally isothermal by setting the pressure to $p=c_s \Sigma(r)$, where $\Sigma(r)$ 
is the surface density (i.e. the vertically integrated density) at radius $r$.
With the above constraints, the 
initial condition for the surface density profile becomes: 
\begin{equation}
    \Sigma(r) = \Sigma_0 \left(\frac{r}{r_0} \right)^{-1/2},
\end{equation}
where $r_0$ is an arbitrary distance unit, and $\Sigma_0$ is a corresponding normalisation.

The simulation domain 
extends from $0.5 < r/r_0 < 6.0$, with 666 logarithmically-spaced radial cells and an
increasing number of azimuthal cells with radius in order to maintain a uniform cell aspect ratio. 
The disc is resolved at all radii to $\Delta r/r_0 \lesssim 0.003$. Fixing the number of cells for all runs means that the resolution in terms of the disc scale height changes with Mach number, such that $\Delta r/h =\lesssim {0.04,0.07,0.1}$ for $\mach = {10,20,30}$, respectively. 
We performed a resolution test with 800 radial cells for our fiducial runs below, and confirmed that the torques did not change at this somewhat higher resolution. 
Compared to Paper I, we more than double 
the outer boundary of the simulation. We found that for more massive satellites, a closer-in outer boundary leads to numerical transients in the torque that last for longer than a viscous time. 

\begin{figure}
\begin{center}
\includegraphics[width=.48\textwidth]{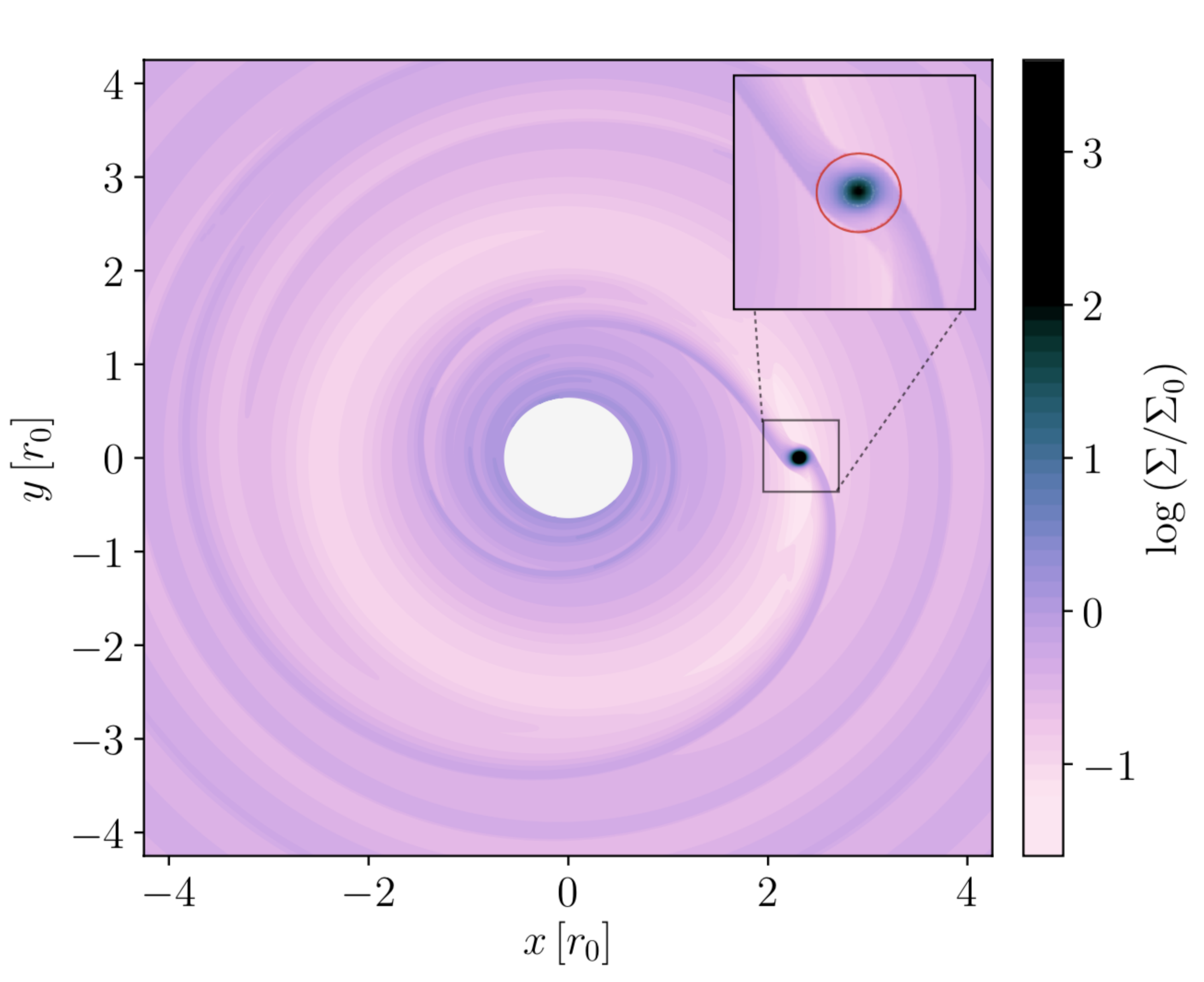}
\caption{Snapshot of the surface density in a simulation with $q=10^{-3}$, $\alpha=0.03$, 
and $\mach=20$ after 2500 orbits of the satellite.  An embedded migrating perturber orbits counter-clockwise, 
excites spiral density waves, and carves a shallow gap.  The zoom-in panel in the upper right corner shows a close-up with the color scaling altered to highlight streams flowing across the secondary's Hill radius (marked by the red circle). 
}
\label{fig:simulation}
\end{center}
\end{figure}

To visualise our simulation, we show a snapshot of the 2D surface density distribution in Fig.~\ref{fig:simulation}.

\subsection{The inspiralling BH}

The primary SMBH is excised from the simulation domain,
and the secondary BH is modeled as a point mass with a smoothed gravitational potential, defined by
\be
\Phi_2 = \frac{G M_2}{({r_2}^2+\epsilon^2)^{1/2}},
\ee
where $G$ is the gravitational constant, $r_2$ is the distance to the secondary BH and $\epsilon = h(r)/2$  is the smoothing length at the secondary's position.
Note that $\epsilon$ is introduced here, as in other similar works, to mimic the vertically integrated forces within a scale height of the BH, and is continuously updated as the BH spirals inward.

Under the quadrupole 
approximation for gravitational radiation, 
for a circular GW inspiral (\citealt{Peters64}), 
the evolution of the separation of a binary driven together by GW emission follows
\be
\label{eq:rdotgw}
\dot{r}_{\rm GW} = -\frac{64}{5} \frac{G^3}{c^5} \frac{M^3}{(1+1/q)(1+q)} \frac{1}{r^3},
\ee
where $M=M_1+M_2$ is the total mass of the binary. 
Following this, the secondary BH is placed in a quasicircular, prograde orbit whose separation evolves according to
\be
r(t) = r_{\rm f} (1 - 4 R_{\rm f}(t-t_{\rm tot}))^{1/4},
\ee
where $t$ is the elapsed time, $t_{\rm tot}$ is the total simulation time, $r_{\rm f}$ is the final binary separation, and 
$R_{\rm f} \equiv -(\dot{r}_{\rm GW}/r)|_{r_{\rm f}}$ is a scaled inspiral rate evaluated at the final separation.

The parameter $R$ can also be considered an inverse residence time, which relates to the time spent at each orbital radius during a GW-driven decay. 
It is useful to express this quantity in terms of the orbital frequency, which provides a dimensionless inspiral rate:
\be
\label{eq:Romega}
\frac{R}{\Omega} = 
\frac{\dot{r}}{\Omega r} = \frac{64}{20 \sqrt{2}} \frac{1}{(1+1/q)(1+q)} \left(\frac{r_{\rm S}(M_1)}{r}\right)^{5/2},
\ee
where $r_{\rm S} = 2 G M_1/c^2$ is the Schwarzschild radius of the primary SMBH and $\Omega$ is the orbital angular frequency.
Note that $\Omega/ R$ is approximately the number of orbits a binary spends at each separation $r$ (the total number of orbits from $r$ to $0$ is $n_{\rm orb}=1/(5\pi) \Omega/R$).

In general, for each simulation, we need to specify an initial and a final position $[r_i,r_f]$ for the inspiraling satellite BH.   In practice, this choice is guided by two considerations.  First, we wish to simulate E/IMRIs that are in the LISA frequency band, close to merger.   Second, while ideally we would follow the inspiral for the entire duration of a LISA observation (4 years by default; see below), in practice, we are limited by numerical considerations to a fixed number of orbits ($N_{\rm orb}=5000$ unless specified otherwise).

The parameterisation above allows us to conveniently implement these choices, by specifying values for $R_{\rm f}$ and $t_{\rm tot}$.  These translate to ranges of $[r_i,r_f]$ which depend on $q$ and the total binary mass.  Motivated by modeling potential LISA sources, we impose inspiral rates $R_{\rm f}$ corresponding to the final stages of a $q=10^{-3}$, $M_1=10^6 M_{\odot}$ fiducial IMRI, reaching $r_f = 3r_{\rm S}$.
Each of our simulations spans $n_{\rm orb} = 5,000$ binary orbits. 
 Provided a final separation $r_{\rm f}$ and a total number of orbits $n_{\rm orb}$, one can solve for the initial separation of the binary through the relation 
\be 
\label{eq:Norb}
N_{\rm orb} = \frac{1}{2 \pi} \int_{r_{\rm i}}^{r_{\rm f}} \frac{\Omega}{\dot{r}_{\rm GW}} dr.
\ee
 With our fiducial parameters listed above, our initial binary separation is $r_i = 8 r_{\rm S}$.

\begin{figure}
\begin{center}
\includegraphics[width=.5\textwidth]{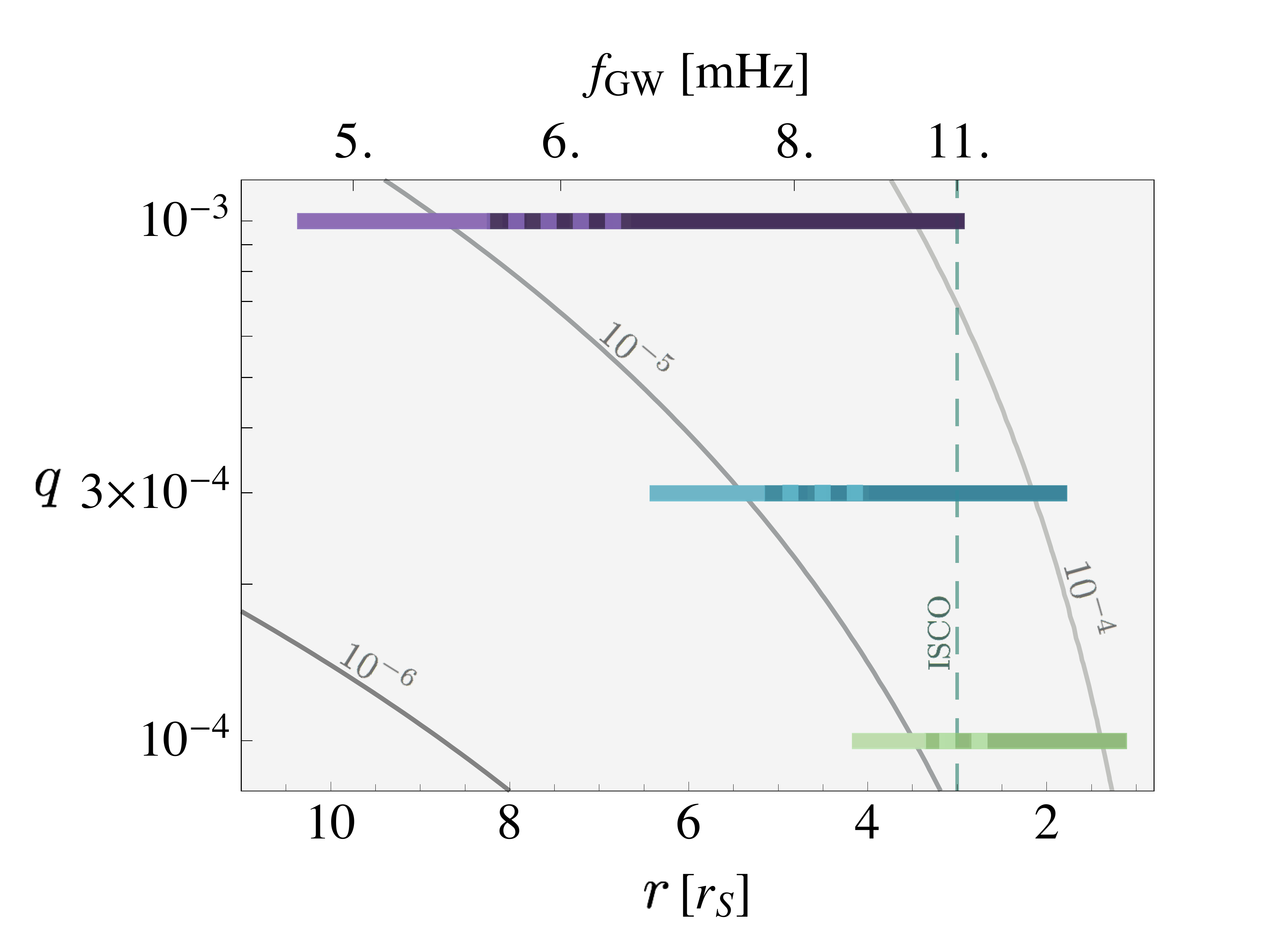}
\caption{Contour lines (in grey) of constant $\dot{r}/(\Omega r)$, a dimensionless inspiral rate, as a 
function of $q$ and binary separation $r$ (or corresponding gravitational wave frequency $f_{\rm GW}$, shown on the upper x-axis
for a fiducial primary mass $M_1=10^6 M_{\odot}$). The horizontal colored lines show the ranges of inspiral rates covered in six of our simulations.  For each mass ratio, these correspond to different separations in Schwarzschild radii. Because we chose to simulate all three mass ratios over the same dimensionless inspiral rate, portions of the two lower mass ratio simulations correspond to separations inside the ISCO (vertical dashed line).}
\label{fig:adot_a_q}
\end{center}
\end{figure}

 While the simulation is scale-free
 (i.e. $r_0=\Sigma_0=1$ in code units),
 prescribing an inspiral rate with Eq.~\ref{eq:rdotgw} implies a physical length scale (i.e. a physical value for $r_0$, or equivalently a value of  $r_0$ in Schwarzschild units). 
We illustrate this conversion in Figure~\ref{fig:adot_a_q},  where we show curves of constant
$\dot{r}/\Omega r$ as a function of $q$ and separation $r$ (or corresponding GW frequency $f_{\rm GW}$ shown on the upper x-axis). We delineate the simulated ranges of binary separations (or equivalently frequency ranges) with straight horizontal lines. Light and dark lines correspond to six different simulations we performed, each spanning 5,000 orbits and probing various physical regimes of coalescence. The dashed portion of the lines indicates overlap, where two different simulations probe the same inspiral rates.

As shown in Fig.~\ref{fig:adot_a_q}, simulations with different $q$ cover the same range of dimensionless inspiral rates ($7 \times 10^{-6} \leq R_{\rm f}\leq 2\times 10^{-4})$, 
despite the 
fact that when decreasing the mass ratio, such rates correspond to unphysically small separations inside the innermost stable circular orbit (ISCO). 
While these inspiral rates are not realised in nature for low-$q$ binaries, 
this academic exercise allows us to isolate the effect of the inspiral on the 
torque, changing one parameter at a time, since we expect torques to be sensitive to the inspiral speed in addition to mass ratio.

In addition to the fiducial simulation that starts at $8 r_{\rm S}$ and reaches the ISCO at $3 r_{\rm S}$ ,
we ran a set of "slow" simulations (lighter lines in Fig.~\ref{fig:adot_a_q}) which probe the $q=10^{-3}$ inspiral from $10 r_{\rm S}$ to $6.5 r_{\rm S}$, similar to the range covered in Paper I.
Rather than simulating a full 10,000 orbits, we split our simulations into two runs that overlap in inspiral rate. 
This provides a sanity check that our results are physical, and not dependent on transients introduced by the initial conditions (in other words, we can test whether the end of the "slow run" yields the same torques as the beginning of the "fast" run).
 Moreover, this reduces the span of radii covered in each single simulation, allowing us to avoid the BH starting too close to the outer boundary.
Each simulation exceeds a viscous time for the disc, which for reference we define as a function of orbital time $t_{\rm orb}$ at the BH position as 
\be
\label{eq:tvisc}
t_{\rm visc} = \frac{2}{3}\frac{r^2}{\nu} \approx 1415 \left(\frac{\mach}{20} \right)^2 \left(\frac{\alpha}{0.03} \right)^{-1} \rm \, t_{\rm orb}.
\ee 

We neglect relativistic effects and keep the potential Newtonian, despite the fact that we are 
simulating regions close to the ISCO, where relativistic effects will affect the dynamics. 
This choice was made primarily for simplicity and to maintain the scale-free nature of the simulation. 

We compute the torques exerted by the gas on the secondary BH, but we neglect their effect on the BH's orbital evolution.
This assumption is justified 
in the regime where the disc mass is insignificant compared to the mass of the secondary BH, and when the torque due to GW emission is far dominant. We demonstrate in \S\ref{sec:results} below that both of these criteria hold to high accuracy. 
This approach keeps the equations scale invariant and $\Sigma_0$ arbitrary, significantly reducing our computational costs and allowing us to run a full parameter study (see \S\ref{sec:suite}).

\subsection{Sink prescription}

Accretion onto the secondary is implemented with the same approach as in \citet{Derdzinski2019}.
The gas inside the smoothing radius $\epsilon$ is approximated as a mini-disc with the same
$\alpha$ and $\mach$ as the global disc. The surface density within a distance $\epsilon$ of $M_2$ 
is decreased on the viscous timescale, 
$t_{\rm visc}(\epsilon)$ 
at a rate
\be
\label{eq:sink}
\frac{d \Sigma}{dt} = -\frac{\Sigma}{t_{\rm visc}(\epsilon)} \exp\left\{-(r_2/\epsilon)^4\right\}.
\ee

As discussed in Paper I, 
the chosen parameters for a thin-disc accretion model correspond to an accretion rate that can be well in excess of the Eddington rate, depending on the normalisation of the disk surface density. 
Gas within the sink radius is depleted at a rate $\dot{M} \approx \pi \epsilon^2 \Sigma_{\rm sink}/t_{\rm visc}$, where $\Sigma_{\rm sink}\! \sim\! 10^3 \Sigma_0$ due to the stark increase in density of gas near the BH for $q=10^{-3}$  (Fig.~\ref{fig:simulation}).
Converting this to a physical rate with $M_2=10^3 M_{\odot}$, $r_0=3r_{\rm S}$, $\mach=20$, and  $\alpha=0.03$ then implies an accretion rate $\dot{M}\gtrsim\dot{M}_{\rm Edd}$ for $\Sigma_0\gtrsim10 \rm \, g \, cm^{-2}$, given $\dot{M}_{\rm Edd}=4 \pi G M_2 / (\epsilon_{\rm eff} \kappa_{\rm es} c)$, where $\kappa_{\rm es}=0.4 \, \rm cm^2 \, g^{-1}$ 
is the electron scattering opacity and we assume a radiative efficiency $\epsilon_{\rm eff}=0.1$. This is a common shortcoming for many accretion disc simulations; it is computationally challenging to simulate highly supersonic discs, and thus we are often limited to unrealistically low Mach numbers.

For an additional reference, if we compare the prescribed accretion rate to a Bondi rate  $\dot{M}_{\rm Bondi} = 4 \pi G^2 M_2^2 \rho_0 /c_s^3$ (e.g. \citealt{Edgar2004}) assuming the relative velocity between the BH and gas is zero (as it is comoving), we find that the accretion rate is well below the Bondi rate ($\dot{M}_{\rm sink}/\dot{M}_{\rm Bondi}\sim10^{-6}$) for any choice of $\Sigma_0$. This suggests that depending on how the accretion flow actually occurs, even higher rates may be feasible. 

Nevertheless, we consider our chosen accretion prescription to be  relatively ``slow'' when compared to others in the literature that deplete gas within the sink on $\sim$orbital timescales. In our case, gas is only partially depleted within the sink radius and only for the highest simulated mass ratio $q=10^{-3}$.  Unlike for more nearly equal-mass binaries (such as those studied in \citealt{Duffell2020}), we find here that
the sink prescription affects the torques for a sufficiently massive gap-opening satellite. We  discuss this further in Section~\ref{sec:results}.

\subsection{Torque measurement}
\label{sec:torqdefs}

Here we describe how the gas disc torque is computed in the simulations and define other torques for comparison. 

When comparing torques on binaries of different mass ratios, it is 
useful to utilize the Type~I formula from \citet{Tanaka2002}, 
\be
\label{eq:T0}
T_0 = \Sigma(r) r^4 \Omega^2 q^2 \mach^2,
\ee
where $\Sigma(r)$ is the initial surface density profile. 
As the BH migrates and the disc evolves from its initial conditions, it is useful to normalise the measured torque by $T_0$ in order to scale out 
the expected radial dependence of the torque. 
Analysing a dimensionless torque allows us to compare results for different mass ratios and isolate the effect of the inspiral.

The dominant mechanism for angular momentum loss on the secondary
 is the torque due to GW emission, 
\be
\label{eq:Tgw}
T_{\rm GW} = \frac{1}{2} M_2 r \dot{r}_{\rm GW} \Omega.
\ee
Note that this is the torque on only one component of the binary.

The primary focus of this work is the gravitational torque 
(also referred to as migration or gas torque)
$T_{\rm g}$, which arises from the gravitational force exerted by the gas on the secondary BH.
We calculate this torque by summing up the $\phi-$component of the gravitational
force ${\bf g}_{\phi}$ crossed with the binary lever arm ${\bf r}$
over all the grid cells in the disc,
\be
\label{eq:Tg}
T_{\rm g} =
\sum | {\bf g}_{\phi} \times {\bf r} |
\ee
where $|\bf r|$ is the distance from the secondary to the center of mass of the binary, which in our setup (and in the limit of $q\ll 1$) is at the origin.  We also discuss the torque density, 
defined by $\mathcal{T}\equiv {\bf g}_{\phi}\times  {\bf r}$ before summation, to analyse 
contributions from different patches of the disc
to the total torque.

Torque can also be imparted via accretion of gas that 
directly adds both 
mass and momentum to the secondary BH (sometimes called accretion torque).
For our parameters we find this torque to be significantly weaker than  the gravitational torque (by several orders of magnitude, as in \citet{Derdzinski2019}, 
so we refrain from discussing it further in the present work.

The magnitude of gas torques
(Eqs.~\ref{eq:T0} ~and~\ref{eq:Tg}) all depend
linearly on the normalisation of the surface density, which is arbitrary
in our simulation setup. Rather than choosing a single value for $\Sigma_0$, we will use this freedom of normalisation to ask: at what value of the surface density do gas torques produce a detectable phase drift in LISA's GW measurements? 
Additionally, is this density physically reasonable, given the densities expected in accretion disc models?

Estimates for accretion disc densities 
in the vicinity of the central SMBH 
vary by several orders of
magnitude.
We adopt two limiting estimates for the surface density normalisation $\Sigma_0$ 
(at the secondary's final radius, taken to be $r_{\rm f}=3 r_{\rm S}$ 
for our detectability estimates below) for the inner regions of thin, near-Eddington
accretion discs.  
They depend on the presumed accretion rate $\dot{M}$ and the viscosity
parameter $\alpha$. The models differ in whether the viscosity scales
with the gas pressure or with the total (gas + radiation) pressure, a choice that has a large
impact when relating the accretion rate to $\Sigma_0$.
Lower viscosity discs require much higher surface densities to maintain the same accretion rate.

We normalise our disc densities to represent AGN accreting at $10\%$ of the Eddington rate
with a radiative efficiency of 10\%.
The first estimate is obtained from the seminal model for a thin,
viscous accretion disc (i.e. $\alpha$-disc;
\citealt{SS1973}). In the inner, radiation-pressure dominated region,
\be
  \label{eq:Sigma_alpha}
\Sigma_{\alpha} = 41.08 \,  \left(
\frac{\alpha}{0.03} \right)^{-1} \! \left( \frac{\dot{M}}{0.1
  \dot{M}_{\rm Edd}} \right)^{-1} \!\left( \frac{r}{3 r_{\rm S}}
\right)^{3/2}
\, {\rm g \, cm^{-2}}.
\ee
In case the viscosity is proportional only to the gas
pressure (i.e. for a so-called $\beta$-disc), the surface density at
the same accretion rate is several orders of magnitude higher,
\begin{multline}
  \label{eq:Sigma_beta}
  \Sigma_{\beta} = 2.11 \times 10^7 \, \left( \frac{\alpha}{0.03} \right)^{-4/5} \! \left( \frac{\dot{M}}{0.1 \dot{M}_{\rm Edd}} \right)^{3/5} \times \\
\times \left( \frac{M}{10^6 M_{\odot}} \right)^{1/5} \! \left( \frac{r}{3 r_{\rm S}} \right)^{-3/5}
\,{\rm g \, cm^{-2}},
\end{multline} 
see \citet{Haiman+2009}.
While both of these models carry a different radial density scaling than our 
disc model, the values are meant to provide a reference for possible 
surface densities, which becomes important for our detectability estimates in \S\ref{sec:LISA} below.
Note that in both these estimates the total disc mass inside the BH's orbit 
is much less than the mass of the BHs.
For example, even with the high-end estimate, an integral of the total enclosed mass within $10 r_{\rm S}$ yields a mass as low as
\be
M_{\rm encl} = 2 \pi \int_{3r_{\rm S}}^{10 r_{\rm S}} \Sigma_{\beta} r dr 
\simeq 0.16 M_{\odot}
\ee
for $\alpha=0.03$, $M=10^6 M_{\odot}$, and $\dot{M} = 0.1 \dot{M}_{\rm Edd}$.

\subsection{Simulation Suite}\label{sec:suite}

In Paper I, we performed a simulation for a single system with ${q,\alpha,\mach}={10^{-3},0.03,20}$, over a range of dimensionless inspiral rates between $7\times10^{-6}$ and $2\times10^{-5}$.
We extend on that study by
expanding the range of inspiral rates to higher values up to $\dot{r}/\Omega r = 2 \times 10^{-4}$, 
and then varying $q$, $\mach$, and $\alpha$. Our fiducial system is a $q=10^{-3}$ binary embedded in a disc 
with $\alpha = 0.03$ and $\mach = 20$. 
We then run simulations for three different mass ratios $q={10^{-3},3\times10^{-4},10^{-4}}$, each 
with three different values of $\alpha = {0.01,0.03,0.1}$. We also run simulations with a range of Mach numbers $\mach={10,20,30}$ 
around our fiducial model.  
In total, we present 15 different simulations, which are listed in Table~\ref{table:parameters}, where each run is 
labeled with its mass ratio and viscosity (for example, `q1e3a03' corresponds to a run with $q=10^{-3}$ and $\alpha=0.03$.)  

For computational feasibility, our study is limited to
higher values of $\alpha$ and lower Mach numbers, in order to avoid prohibitively long viscous times to reach steady state (see Eq.~\ref{eq:tvisc}).
Our fiducial choice of $\mach=20$ corresponds to a much hotter and thicker disc than expected in thin, near-Eddington AGN discs, where continuum emission suggests Mach numbers exceeding $\mach\sim 100$ \citealt{Krolik1999}).
However, higher Mach numbers are 
numerically challenging to simulate, primarily because highly supersonic flows require increasingly high resolution as they develop complicated, small scale features.
As we show in \S\ref{sec:mach} below, increasing the 
Mach number to $30$ already produces
very noisy torques and a dense and unstable gas morphology.

\begin{table*}
\begin{tabular}{llllc|l}
\multicolumn{6}{c}{ \sc Simulations} \\
\specialrule{0.8pt}{1pt}{1pt}
\vspace{-0.3cm} \\
\multicolumn{1}{l}{ Name} &
\multicolumn{1}{l}{ Mass Ratio} &
\multicolumn{1}{c}{ Viscosity} &
\multicolumn{1}{c}{ Mach Number} &
\multicolumn{1}{c}{ Separation $[r_{\rm S}]$} &
\multicolumn{1}{|c}{ Average Torque} 
\\
\midrule
\vspace{-0.4cm} \\ 
\multicolumn{1}{l}{ } &
\multicolumn{1}{l}{ $q$} &
\multicolumn{1}{l}{  $\alpha$ } &
\multicolumn{1}{l}{ $\mach$} &
\multicolumn{1}{c}{ $[r_i,r_f]$} &
\multicolumn{1}{|c}{  $\langle T_{\rm g}/T_0\rangle$} 
\\
\midrule
q1e3a03   &  $10^{-3}$&  $0.03$   &  $20$   & $ [10.3, 6.5] $  & $0.19$ \\
 (fiducial)  & & & &  $[8.2,3.0]$  & *see Fig.~\ref{fig:Ttotal_adot} \\
   \specialrule{0.2pt}{1pt}{1pt}
q3e4a03   &  $3\times10^{-4}$    &  $0.03$   &    $20$   & $[6.4,4.0]$   & $0.01$ \\ 
   & & & & $[5.1,1.9]$  &  *see Fig.~\ref{fig:Ttotal_adot}\\
   \specialrule{0.2pt}{1pt}{1pt}
q1e4a03   &   $10^{-4}$   &  $0.03$ & $20$ & $[4.1,2.6]$ & $-0.29$\\  
   & & & & $[3.3,1.2]$  &  *see Fig.~\ref{fig:Ttotal_adot}\\
   \specialrule{0.2pt}{1pt}{1pt}
q1e3a1   &  $10^{-3}$   &  $0.1$   & $20$ &  $[8.2,3.0]$ & $0.34$\\ 
   \specialrule{0.2pt}{1pt}{1pt}
q3e4a1  &  $3\times10^{-4}$   &  $0.1$ & $20$ &  $[5.1,1.9]$ & $0.38$ \\ 
   \specialrule{0.2pt}{1pt}{1pt}
q1e4a1  &  $10^{-4}$  &  $0.1$ & $20$  & $[3.3,1.2]$  & $-1.26$\\ 
 \specialrule{0.2pt}{1pt}{1pt}
 q1e3a01   &  $10^{-3}$   &  $0.01$   & $20$ &   $[8.2,3.0]$ & $\sim\,0.01$ (*see Fig.~\ref{fig:Ttotal_adot_a0p1})\\ 
   \specialrule{0.2pt}{1pt}{1pt}
q3e4a01  &  $3\times10^{-4}$   &  $0.01$ & $20$ &  $[5.1,1.9]$ & $\sim\!-0.51$ (*see Fig.~\ref{fig:Ttotal_adot_a0p1}) \\ 
   \specialrule{0.2pt}{1pt}{1pt}
q1e4a01  &  $10^{-4}$  &  $0.01$ & $20$  & $[3.3,1.2]$  & $\sim\!-0.25$ (*see Fig.~\ref{fig:Ttotal_adot_a0p1}) \\ 
 \specialrule{0.2pt}{1pt}{1pt}
q1e3a03m10 &  $10^{-3}$ & $0.03$ &$10$  & $[8.2,3.0]$ & $-0.30$  \\ 
\specialrule{0.2pt}{1pt}{1pt}
q1e3a03m30 & $10^{-3}$ & $0.03$ & $30$  & $[8.2,3.0]$ & $\sim0.03$ (*see Fig.~\ref{fig:Ttotal_machs})\\
\specialrule{0.2pt}{1pt}{1pt}
q1e3a03 (no sink)   &  $10^{-3}$&  $0.03$   &  $20$   & $ [8.2, 3.0] $  &  *see Fig.~\ref{fig:Thill}\\

\specialrule{0.8pt}{1pt}{1pt}
\end{tabular}
\caption{List of our 15 simulations and their parameters. Each binary 
is forced to inspiral at the GW-driven rate from its initial $r_i$ to its final separation $r_f$ (listed in Schwarzschild units). In those cases where we ran two simulations for a single binary to probe different inspiral rates, two ranges of radii are listed (first three rows).
We also show the average torque value (measured over the last 2000 binary orbits) in simulations
for which we calculate the SNR of the gas-induced deviations in the GW waveform. *Depending on the disc and system parameters, some simulations show deviations from the average torque throughout the GW inspiral -  these we denote in the last column with a reference to the corresponding Figure. }
\label{table:parameters}
\end{table*}

\section{Results}
\label{sec:results}

Here we describe results from our simulations -- measuring and analysing the torques exerted 
on the satellite BHs -- before deriving estimates of the detectability of the corresponding imprints in the LISA GW waveforms.

\subsection{Gas torques depend on parameters}
\label{torqresults}

As predicted by analytical estimates such as $T_0(r,q,\mach)$, we expect that 
torques will not only depend on system parameters but also evolve with 
radius as the binary separation decreases. 
We find that the \emph{magnitude} of gas torques generally agrees, within an order of magnitude, with the analytical predictions $T_0(q,\mach,r)$ (Eq.~\ref{eq:T0}). However, the \emph{direction} of the torque is difficult to predict. 
Torques tend to oscillate around a value close to (typically less than) $T_0$, but at the late stages of the inspiral, the evolution may deviate from the expected scaling. 
As we discuss in the following sections, 
whether torques are negative (inward) or positive (outward) depends on the combination of 
$q$, $\alpha$, and $\mach$, and we find there is no direct or obvious scaling in this intermediate mass ratio regime.

\subsubsection{Mass ratio}
\begin{figure}
\begin{center}
\includegraphics[width=.5\textwidth]{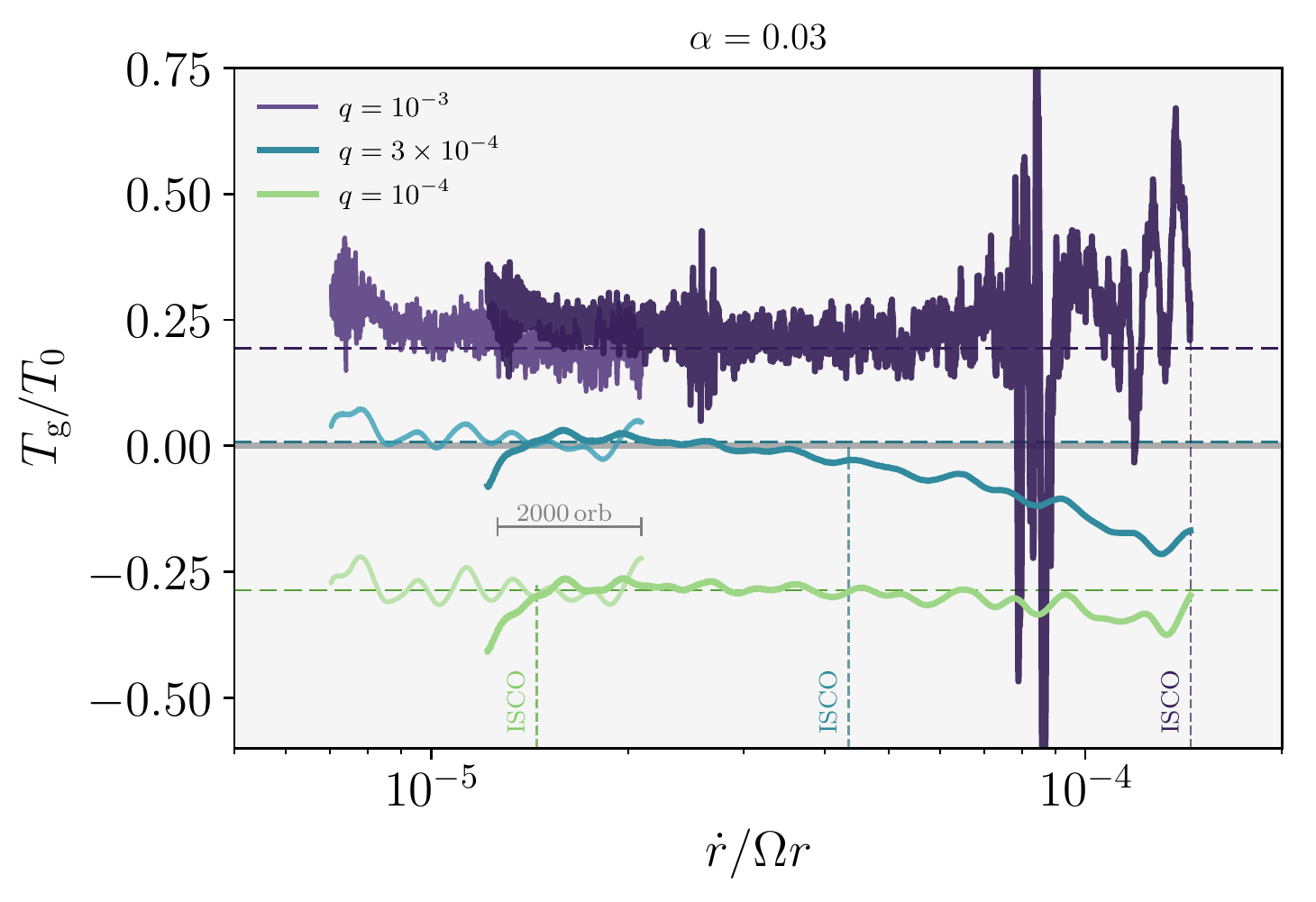}
\caption{Gas torque (normalized by the Type~I torque $T_0$) as a function of dimensionless inspiral rate.
Each color corresponds to a different mass ratio, and each mass ratio is covered by two simulations overlapping in the inspiral rates (lighter lines on the left, corresponding to slower inspiral, and darker on the right, corresponding to faster inspiral). 
Dashed horizontal lines show the average normalised torque, measured over the last 2000 orbits of the slower inspiral runs, as marked by the grey bracket. 
Interestingly, the absolute values of the gas torque are similar in the $q=10^{-4}$ (lowest curves) and $q=10^{-3}$ (uppermost curves), while their signs differ; this appears to be a coincidence.
For simulations reaching fast inspiral rates, the torque begins to deviate from the average for the two higher mass ratios. Vertical dashed lines mark which inspiral rate corresponds to the ISCO for each $q$.}
\label{fig:Ttotal_adot}
\end{center}
\end{figure}

In Fig.~\ref{fig:Ttotal_adot}, we plot the measured $T_g/T_0$ as a function of inspiral rate (Eq.~\ref{eq:Romega}, 
itself a function of $r$)
 measured in simulations with
different mass ratios in a disc with fiducial parameters $\alpha=0.03$ and $\mach=20$.  
As we expect from similar studies of torques on stationary (i.e. non-migrating) satellites,  
torques in this regime are sensitive to even small changes in system parameters. 
In fact, when increasing $q$ from $10^{-4}$ to $10^{-3}$,
the figure shows that the direction of the torque changes. This particular behavior is, however, also dependent on the choice of $\alpha$ (see below).

Additionally, the torque exerted on the $q=10^{-3}$ satellite is significantly more noisy compared to the smaller mass ratios. This is attributed to the large pile-up of gas that accumulates very close to the BH in this mass regime. We discuss this behavior and its consequences in \S\ref{sec:torqHill} below.

\begin{figure}
\begin{center}
\includegraphics[width=.49\textwidth]{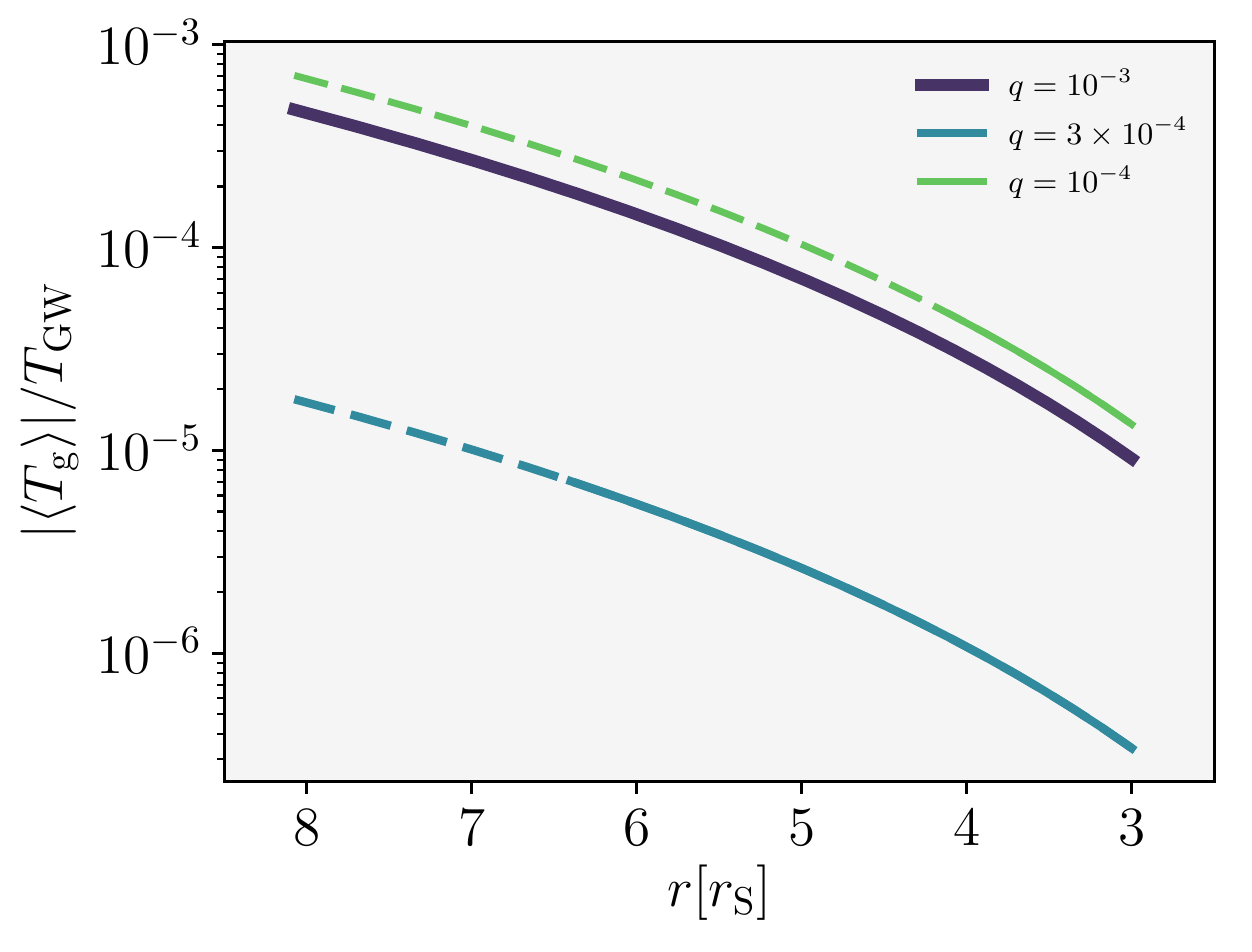}
\caption{Average gas torques from simulations with three different mass ratios with $\alpha = 0.03$ and $\mach=20$ (shown in Fig.~\ref{fig:Ttotal_adot}), scaled by the GW torques during the last few $r_{\rm S}$ prior to ISCO. The dashed portion of each curve indicates the radii for which the torque is extrapolated outside of the simulated range (see Fig.~\ref{fig:adot_a_q}). Gas torques are normalized by $\Sigma_{\beta}$, a high-end estimate for the disc surface density (Eq.~\ref{eq:Sigma_beta}). Gas torques are significantly weaker than GWs in this regime, but can accumulate to a detectable effect over the large number of cycles in the LISA band (see \S\ref{sec:LISA}).}
\label{fig:Tgw}
\end{center}
\end{figure}

First we compare the strength of the gas torque to GWs.
We compute the average torque 
from the last 2000 orbits in each of the slower-inspiral runs
(where the scaled torques $T_g/T_0$ are constant to a good approximation), 
and compare these to the GW torques during the last stages of coalescence. These comparisons are shown in Fig.~\ref{fig:Tgw}. 
The average torque measured 
in the simulation is extrapolated to earlier times for the two smaller mass ratios, assuming that the scaled torque remains constant (shown by the dashed curves).
To scale 
the torques to physical values, we normalise the surface density with the $\beta$-disc estimate (Eq.~\ref{eq:Sigma_beta}), making 
these high-end estimates of the disc torques.
In all cases, despite the high assumed disc density, gas torques are several orders of magnitude 
weaker than that due to GWs at this stage of coalescence. 
However, as we will show in \S\ref{sec:LISA}, over an observation of many thousand cycles 
even weak gas impact can accumulate and produce detectable signatures in the measured GW signal.

In most cases we explore here, the average $T_g$ is within a factor of a few of the analytical estimate $T_0$, 
but deviates towards stronger torques (either negative or positive) as the inspiral rate 
increases. This happens at the final stages of the inspiral (the final $\sim1000$ orbits),
and in the case of the smallest mass ratio $q=10^{-4}$, in reality the merger occurs before the inspiral
has a chance to substantially affect the torque. However, as we describe below, when or whether this 
deviation occurs depends on other disc properties.

\subsubsection{Viscosity}

In this highly nonlinear regime, viscosity affects several important aspects to this system, each of which contribute to the torque in significant ways.  First, viscosity affects the gap depth, which will affect the magnitude of the Lindblad torque excited in the disc.  Secondly, viscosity sets the rate that angular momentum can be carried away from the corotation resonances, and whether corotation torques are in a ``saturated'' or ``desaturated'' state \citep{Masset2006,Duffell2015}.  Third, viscosity can suppress instabilities in the disc that might otherwise generate vortices which exert nontrivial time-dependent torques on the perturber.  Finally, in this complicated nonlinear regime, it is fully possible there are comparable additional torque affects that have not been considered yet, as most analysis of viscous affects on torque are carried out in the linear or weakly nonlinear regimes.  All of this adds up to the result that the torque has a nontrivial $\alpha$ dependence, and even small changes in $\alpha$ 
 (at the least of order $10^{-2}$, given the values explored in the present work)
 
can change the sign of the torque. This is 
 seen in simulations of a range of mass ratios from planetary scales \citep{Duffell2015} up to $q\sim0.5$ in the context of MBH binaries \citep{Duffell2020}.

\begin{figure*}
\begin{center}
\includegraphics[width=.48\textwidth]{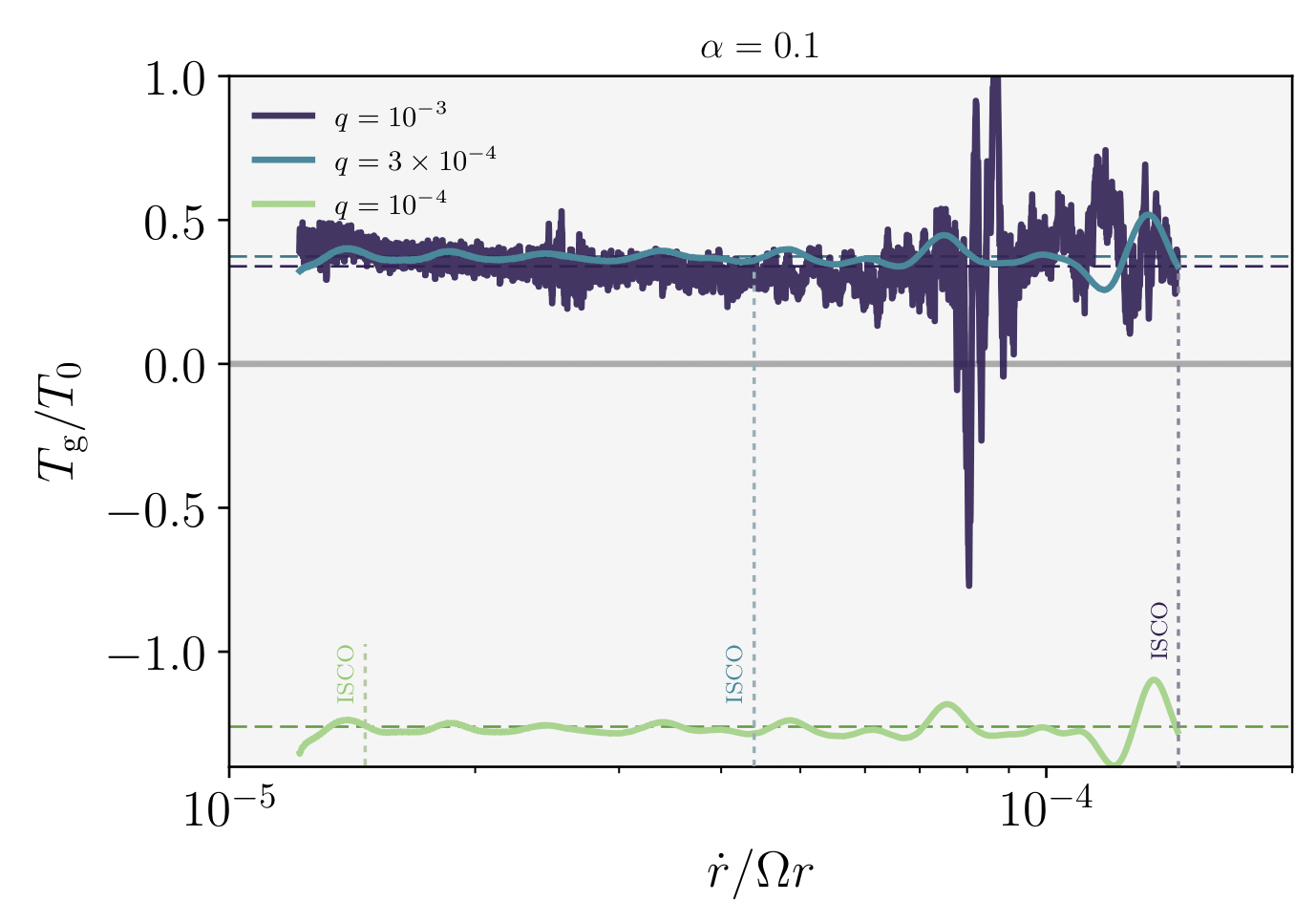}
 \includegraphics[width=.48\textwidth]{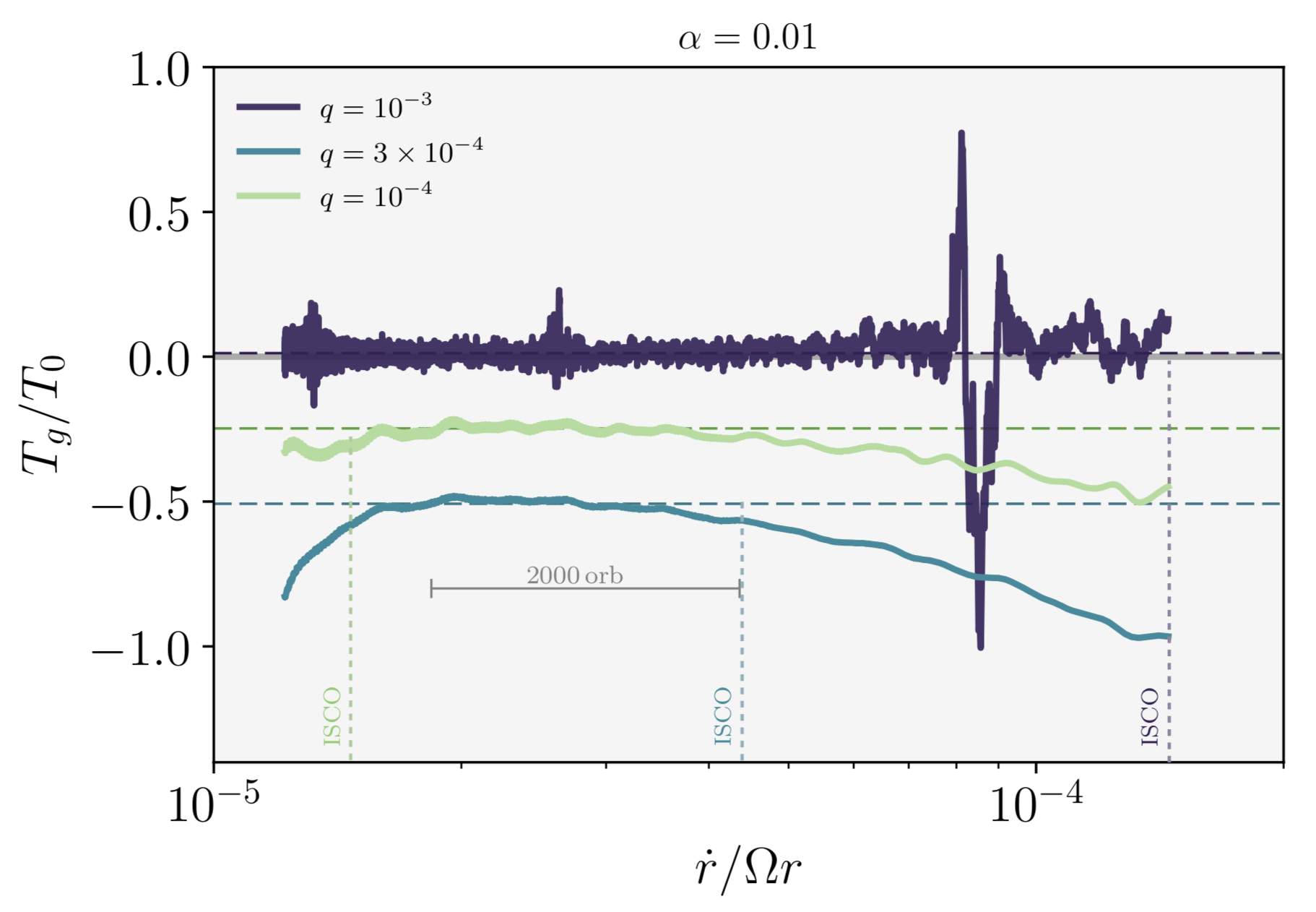}
\caption{\emph{Left panel:} Gas torque normalized by $T_0$ for three different mass ratios, in a disc with $\alpha=0.1$. Higher viscosity means the 
disc stabilises more quickly, and torques do not change as significantly as the inspiral accelerates.
\emph{Right panel:} Gas torque normalized by $T_0$ for runs with $\alpha=0.01$. Dashed horizontal lines 
in both panels
indicate an average taken over a 2000-orbit window depicted by the grey bracket. }
\label{fig:Ttotal_adot_a0p1}
\end{center}
\end{figure*}

We have run additional sets of simulations for each mass ratio in discs with higher and lower viscosities 
($\alpha=0.1$ and $\alpha=0.01$) to observe changes in the 
overall torque and its evolution with the inspiral rate. The results are shown in Fig.~\ref{fig:Ttotal_adot_a0p1}. 
We find that the magnitude $|T_{\rm gas}|$ typically increases in strength with $\alpha$, as has been 
observed in other numerical studies (e.g. \citealt{RobertCrida2018}),
although the direction changes unpredictably,
particularly for $q=3\times10^{-4}$.

Intuitively, we expect that for higher values of viscosity, the inward migration driven by GW emission
should have less impact on the torque, 
simply because the disc can reorganize more quickly 
 (recall that $t_{\rm visc} \propto \alpha^{-1})$, and thus a stronger viscosity leads to a faster radial inflow\footnote{This is further supported by the idea that a higher viscosity will inhibit gap formation, given that more efficient angular momentum transport leads to a rapid replenishing of gas in the co-orbital region 
 \citep{2007astro.ph..1485A}.}).  
Indeed, this is what we observe in the top panel of Fig.~\ref{fig:Ttotal_adot_a0p1}. 
For $\alpha=0.1$ the torque
is essentially constant throughout the entirety of the inspiral (modulo short time-scale oscillations) for all mass ratios, 
and its absolute value can be approximated by $T_0$ to within a factor of two.
For the weaker viscosity, the lower panel of Fig.~\ref{fig:Ttotal_adot_a0p1} shows that $T_g/T_0$ deviates from the nearly constant value during earlier stages, leading to an increasingly negative torque.

In the case of thin, fully ionised discs in near-Eddington AGN, observational evidence suggests that viscosity may reach values of $\alpha=0.1-0.4$ \citep{KPL2007}. Our results therefore imply that E/IMRIs in such viscous discs may be subject to stronger torques, but are less likely to 
show significant changes in torque during the inspiral. 
However, it is the combination of $\alpha$ and $\mach$ that determines the disc dynamics (note the kinematic viscosity $\nu\propto\alpha/\mach^2$), and the values we consider here provide AGN-like characteristics (e.g. gap depth). 
Furthermore, in reality $\alpha$ may scale with radius - 
thus it is possible that the torque evolution seen in lower $\alpha$ simulations may still be relevant for AGN.
In Section~\ref{sec:LISA}, we discuss the case in which torques follow a simple scaling with radius.

\subsubsection{Mach number}
\label{sec:mach}

The Mach number, a measure of disc temperature and thickness, is also a critical factor in
determining the gap depth and disc dynamics near the BH \citep{Duffell2015}.
A low Mach number disc is subject to
stronger pressure forces, resulting in shallower gaps, while a higher Mach number describes a
dynamically colder disc that can consequently form deeper gaps.
For high enough Mach number the scale height of the disc (recall $h/r=1/\mach$) can approach or fall within the Hill sphere of the perturber, leading to a more dynamic gas flow across the gap (see below).

For our fiducial system with $q=10^{-3}$ and $\alpha=0.03$, we explore a range of Mach numbers 
from $\mach=10-30$.  
While this range is limited,
we are able to observe trends in gap depth and gain insight into the
dependence of the gas dynamics close to the BH on $\mach$, which gets increasingly complex for colder discs. 

\begin{figure*}
\begin{center}
\includegraphics[width=.32\textwidth]{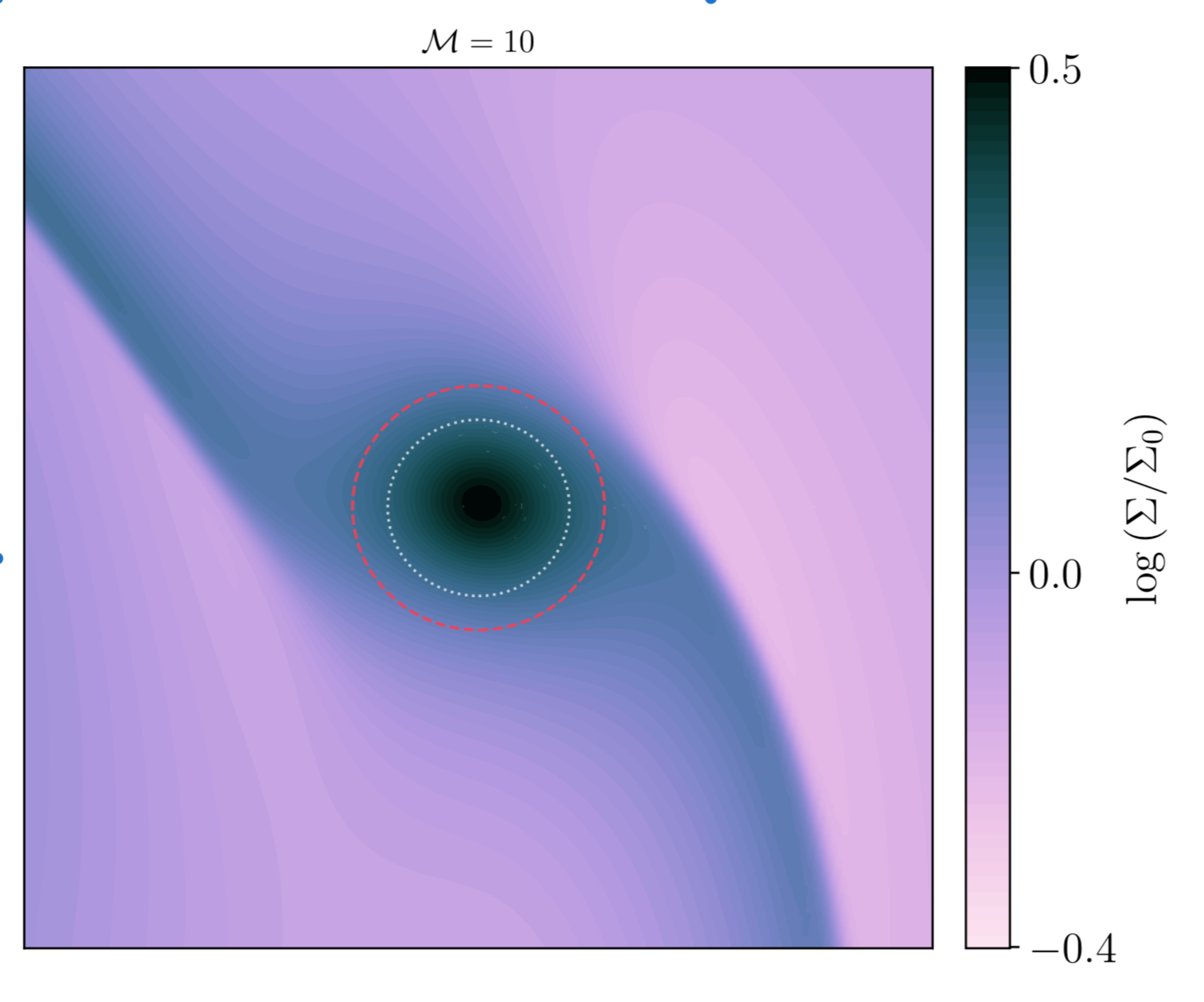}
\includegraphics[width=.32\textwidth]{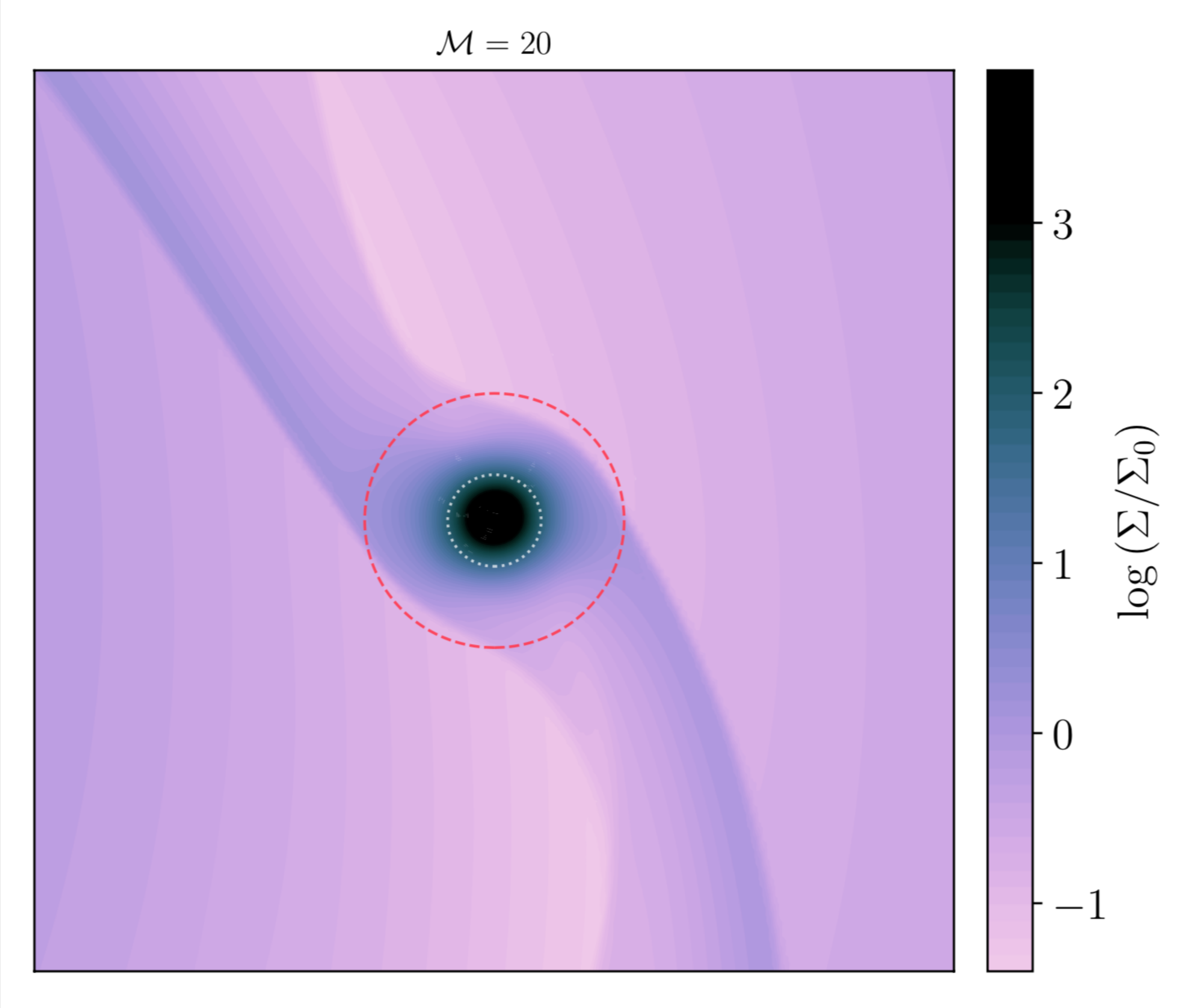}
\includegraphics[width=.32\textwidth]{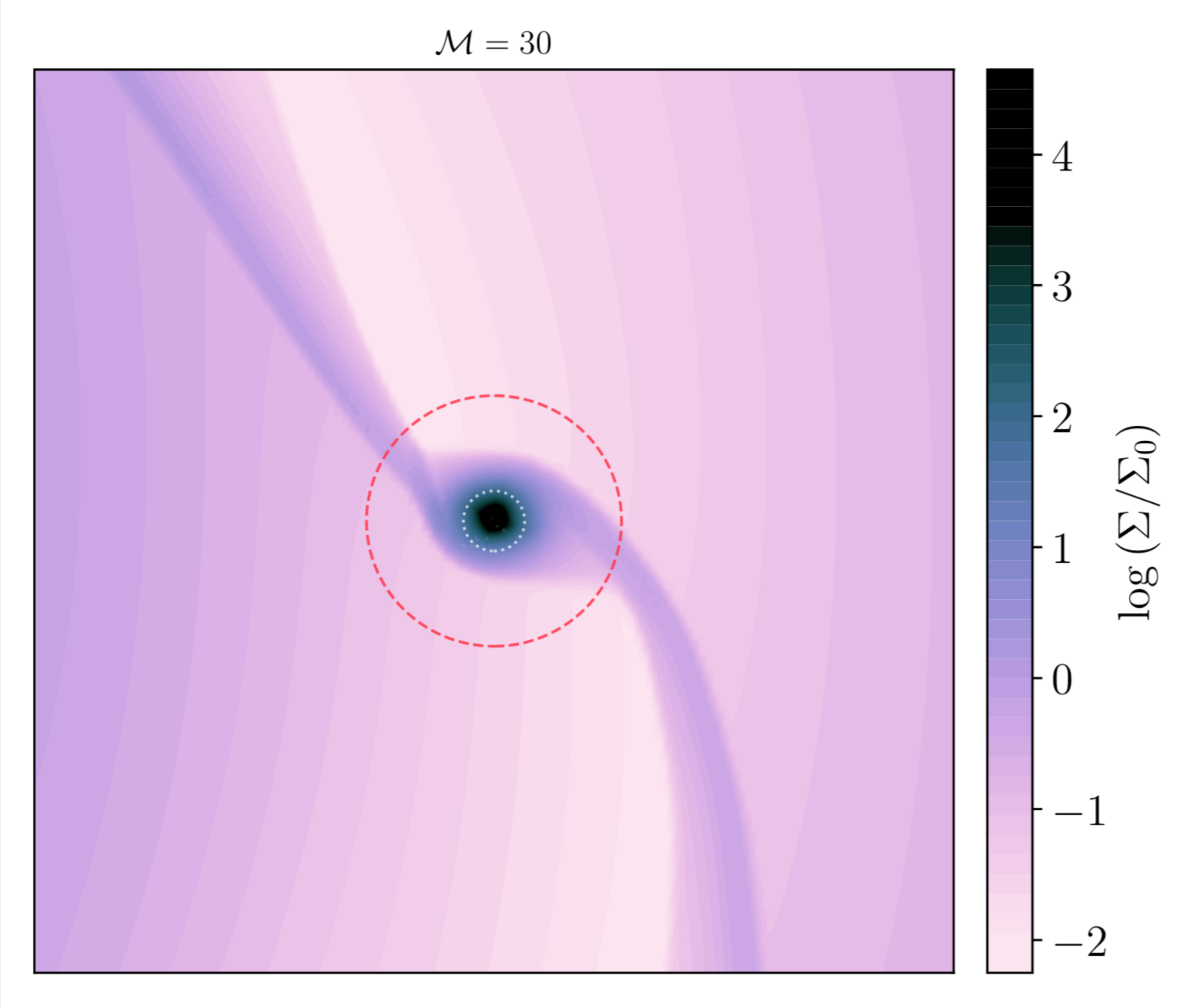}
\caption{Snapshots of the surface density, zooming in on the region close to the BH.  
for three different Mach numbers, as labeled on the top of each panel. For comparison we show the Hill sphere (red circles) and the disc scale height (dashed yellow circles).
Lower Mach 
number discs are hotter and subject to stronger pressure forces that smooth out the flow. For higher 
Mach numbers, as the scale height becomes smaller than the Hill sphere, gas flow across the gap occurs along thinner streams and the BH carves a much deeper gap.
Note the different color scalings in each of the panels;
the surface densities in the $\mach=30$ case are several orders of magnitude higher.}
\label{fig:dens_machs}
\end{center}
\end{figure*}

Fig.~\ref{fig:dens_machs} shows surface density contours of gas close to the 
BH at the end of each simulation. 
For the lowest Mach number ($\mach=10$, which, we note, is a value often adopted in binary simulations), 
pressure forces significantly smooth the flow. 
For the highest Mach number ($\mach=30$), gas flows more tightly around the BH (indeed, 
the scale height and corresponding smoothing length $\epsilon$ is smaller). 
Gas morphology within the Hill radius is dynamic, with narrower streams that flow across a 
deeper gap and stark density contrasts that lead to instabilities in the gap edges and streams.

\begin{figure}
\begin{center}
\includegraphics[width=.5\textwidth]{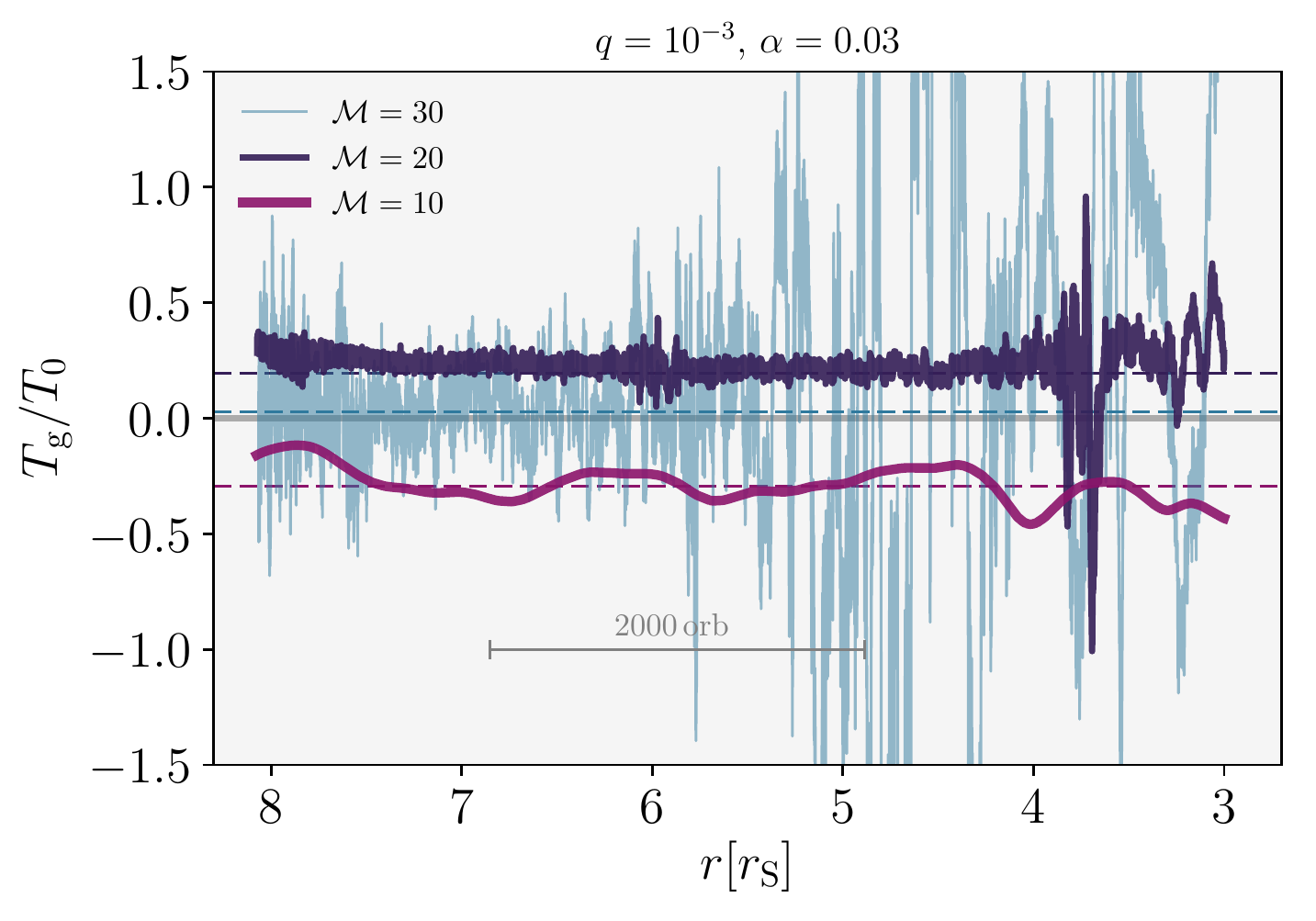}
\caption{Gas torque on a $q=10^{-3}$ inspiral in a disc with $\alpha=0.03$, for three different Mach numbers. Dashed lines show averages taken over the depicted 2000 orbit window.
With higher Mach number, the torque 
shows large variability throughout the inspiral, with amplitude increasing especially at the later stages.}
\label{fig:Ttotal_machs}
\end{center}
\end{figure}

Fig.~\ref{fig:Ttotal_machs} shows the corresponding gas torques.
In the warmer disc ($\mach=10$; purple curve), the torque is smoother and \emph{negative},  rather than positive as in the fiducial $\mach=20$ case (blue curve).  For the colder disc ($\mach=30$; blue curve), the net torque becomes positive, but the most striking feature is that
the dynamic flow around the BH produces a highly variable torque.  The variability increases dramatically as the BH migrates inward.

In order to verify that the large fluctuations are physical and not due to under-resolved gas flow near the BH, we
performed  higher-resolution runs as a test (up to 800 radial cells, corresponding to $\Delta r \gtrsim h/10$). These tests show that the variability persists, but the amplitude and timescale are not yet converged. In fact, a higher resolution leads to larger amplitude in torque fluctuations.
This leads us to believe that the fluctuations are physical, and that highly supersonic discs are indeed more sensitive to the evolution of an embedded BH during a GW inspiral. Without stronger pressure forces to smooth out fluctuations in the flow, gas torques 
become more unstable.
A more detailed investigation of the impact of such a variable torque on a GW inspiral will be investigated in future work.

\subsection{Dissecting the torque}
\label{sec:torqHill}

With our simulations we have the ability to assess contributions of different parts of the disc to 
the net gravitational torque. 
Of particular interest is being able to distinguish between gas very close to the satellite BH, 
i.e. near and within its Hill radius, and that from elsewhere in the disc (the inner and outer discs as 
well as streams crossing the gap). 
One might compare this to corotation torques discussed in the planetary migration literature, keeping in mind that these studies concern gas within the entire co-orbital region rather than just the Hill sphere.
Similar to what we find here,  numerical studies find that corotation torques can become positive in discs with steep density gradients \citep{Paardekooper2008} or with sufficiently high viscosities \citep{Masset2006,Duffell2015}.
An important distinction to note is that corotation torques  
include streams of gas in the horseshoe region, 
 which includes the entire annulus of the migrator and often excludes the Hill radius. 

Here, we do not consider gas within the entire co-orbital annulus of the BH, 
because most of this gas is at low densities due to gap-opening. 
Instead we focus on the gas within the Hill radius of the secondary. In our case, streams in this region that flow across the gap are both dense and close to the BH, and thus dominate the ``corotation'' torque.  As we show, this contribution to the torque arises from a back-to-front asymmetry in the gas morphology at or inside the Hill radius.

The Hill radius, defined as the region where the orbital velocity of gas bound to the satellite BH matches the orbital velocity of the BH itself, is given by
\be
r_{\rm H} = \left( \frac{q}{3} \right)^{1/3} r.
\ee

\begin{figure*}
\begin{center}
\includegraphics[width=.95\textwidth]{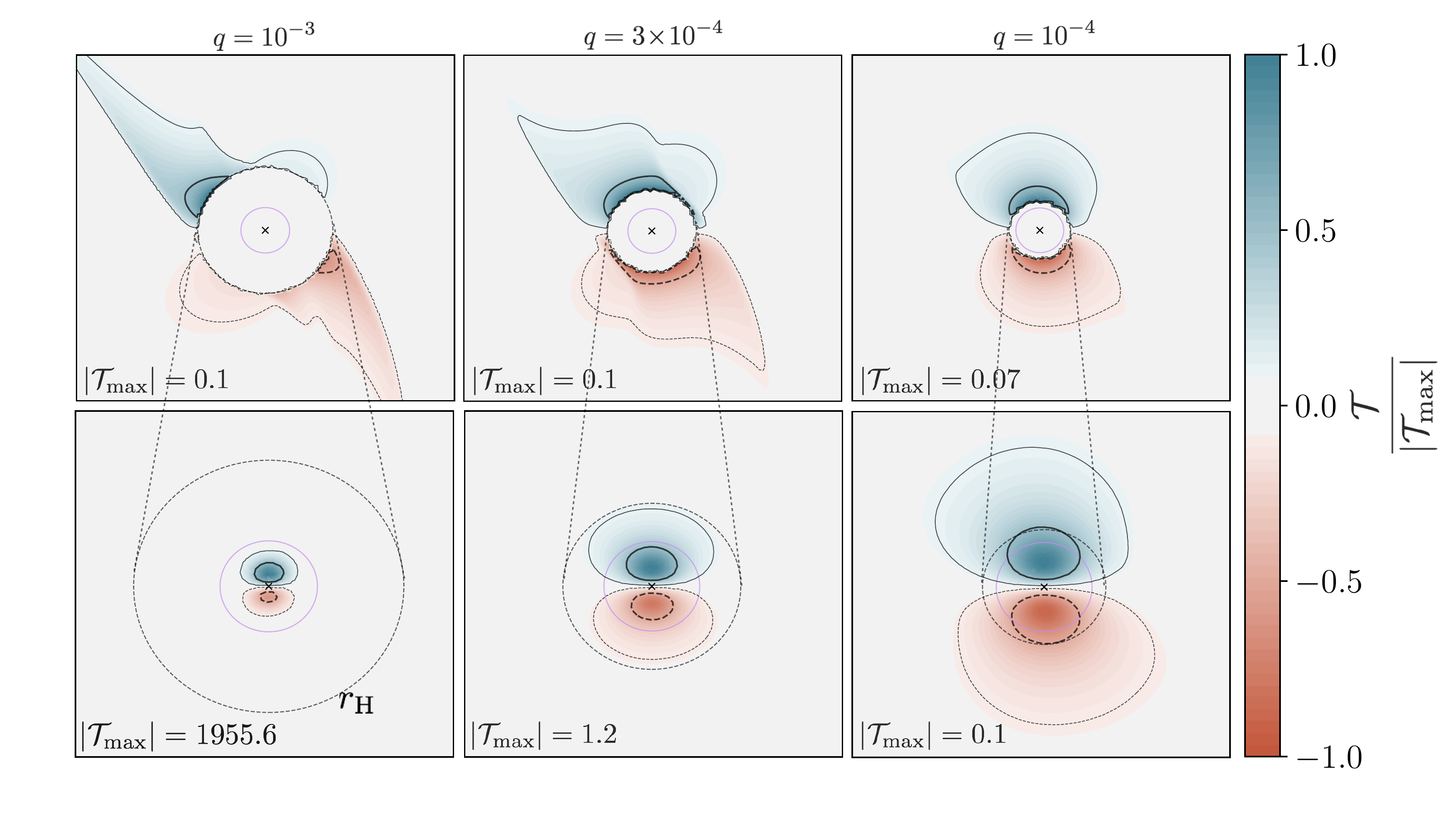}
\caption{2D contours of torque surface density ($\mathcal{T}\equiv  {\bf g}_{\phi}\times  {\bf r}$, 
equivalently a torque per unit disc surface area) close to the satellite BH for discs with $\alpha=0.03$ and $\mach=20$ and three different mass ratios. In the \emph{top panels} we 
have excised the gas in the Hill radius ($r_{\rm H}$; dashed circles) to highlight the gas morphology in streams 
nearby. \emph{The bottom panels} show zoom-in views of the torque contributed by gas within the Hill sphere. 
All contours are normalized by the maximum $\mathcal{T}$, printed in each panel for reference. 
Note that gas pile-up for $q=10^{-3}$ is deep within the Hill radius and reaches significantly high 
densities. This results in much higher torque densities. The smoothing length of the gravitational 
potential is denoted with the solid purple circles.}
\label{fig:torq_grid}
\end{center}
\end{figure*}

As discussed in Paper I already for the $q=10^{-3}$ case, gas pile-up close to the BH becomes non-negligible. 
This can be seen in the density contrasts in 
Fig.~\ref{fig:dens_machs},
or in Fig.~\ref{fig:torq_grid} where we show various 
distributions of torque density $\mathcal{T}$ in our fiducial runs for each $q$. 
Excising the Hill radius (top panels) allows us to observe contributions to the torque from nearby streams, while zooming in on the Hill radius (bottom panels) allows us to analyse the dominant torque contributions  (due to its proximity to the BH).  Most notable is the high torque density for $q=10^{-3}$ arising from the high-density atmosphere that accumulates around the BH. Despite such high values of $\mathcal{T}$,  the net torque in this region remains below $T_0$, implying that as density of the circum-BH gas increases, so does its degree of front-to-back symmetry. 
This is likely a symptom of gas accumulating within a deeper potential well around a more massive secondary BH, which naturally results in a more compact (yet still asymmetric) distribution.

\begin{figure}
\begin{center}
\includegraphics[width=.49\textwidth]{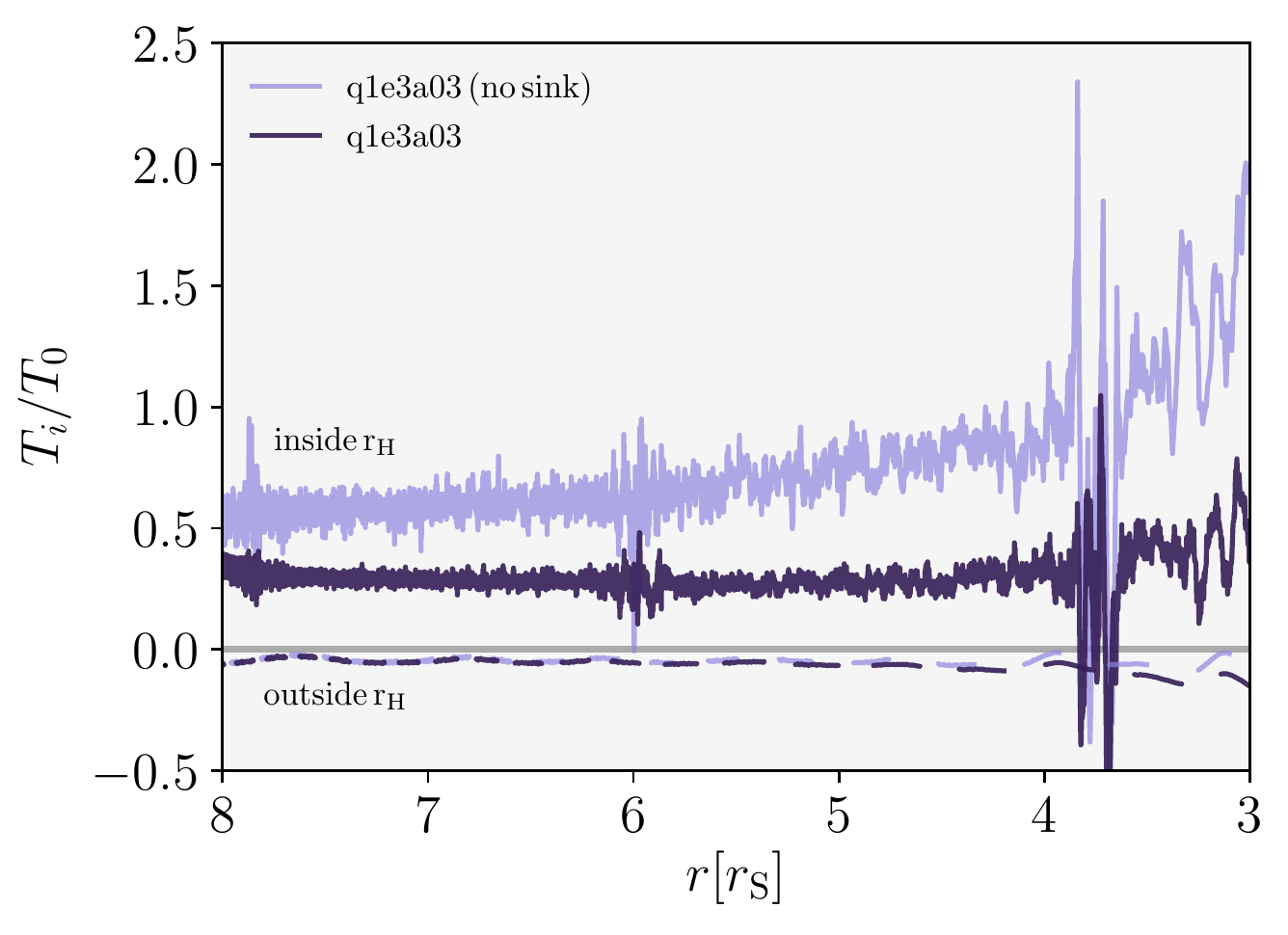}
\caption{Torque on a $q=10^{-3}$ mass ratio inspiral divided into contributions from 
within and outside of the Hill radius, compared to the torques in a run with no accretion
(i.e. with the sink turned off). 
The contributions from inside [outside] $r_{\rm H}$ are positive [negative].
Accretion reduces the gas density and damps the torque component inside the Hill radius, which otherwise would be 
affected by the increasing inspiral rate.}
\label{fig:Thill}
\end{center}
\end{figure}

Gas within the Hill radius contributes a substantial fraction to the total torque, as 
we show in Fig.~\ref{fig:Thill}. This figure shows separately the torque contributions from gas within and 
outside of the Hill radius for the fiducial $q=10^{-3}$ run. Gas inside $r_{\rm H}$ is 
responsible for the positive component of the torque and also shows the strongest dependence on 
the inspiral rate, increasing from $\sim4 r_{\rm S}$ to $3  r_{\rm S}$ (see the solid lines in Fig.~\ref{fig:Thill}). 
This region is sometimes assumed to not contribute to the torque, or is damped by a manual "tapering" function (see, e.g. \citealt{Dempsey2019}), 
even though this gas is a crucial component of material flow across the gap and may exhibit non-negligible asymmetries. Indeed, in simulations by \citet{Crida2009}, the migration rate of a live planet was found to be dependent on how much the torques from the material inside the Hill sphere are damped,  suggesting that this gas is a significant contributor to the net torque, and indeed a physical component that should not be excluded.

In the present work, the Hill sphere torque is of particular importance as the asymmetry 
near the BH may be exacerbated during a sufficiently fast inspiral (here driven by GWs). We expect that 
any changes in the torque during a GW-driven inspiral will first arise from gas closest to the BH. 
This is shown more clearly by comparing versions of our fiducial run for which the sink prescription is turned off (shown as light curves in Fig.~\ref{fig:Thill}). Unsurprisingly, an accreting BH experiences less positive torque
as density within the Hill radius is depleted (while it is nearly unaffected outside).
Without accretion, torques from within $r_{\rm H}$ substantially increases 
with the inspiral rate, more than doubling in comparison to $T_0$ within the final 5,000 
orbits of the inspiral (from $8r_{\rm S}$ to $3 r_{\rm S}$). 
Conversely, we expect that for arbitrarily fast accretion rates (e.g. if the BH manages to continually deplete gas in the Hill region), the positive contribution to the torque would diminish and the BH would instead feel a stronger, negative torque exerted by the rest of the gas in the disk.

For satellites with mass ratios below $q=10^{-3}$ accretion has a negligible 
impact on the torque $-$ when comparing runs with and without accretion, we see no distinct difference in the surface density profile or the torque, given that smaller
satellites accrete an insignificant amount of material according to our sink prescription.  

 The sensitivity to accretion prescription  observed for $q=10^{-3}$ raises the possibility that for IMRIs, detecting torque evolution with frequency may provide insight into the gas dynamics near the BH. 
We hypothesize that this accretion dependence would occur for higher $q\gtrsim 10^{-3}$, although this regime remains to be explored. However, we note that for mass ratios above $q\sim0.04$, the gas dynamics transitions from an annular gap to a circumbinary cavity \citep{DOrazio2016}. In this regime, recent simulations confirm that the sink rate does not affect the torque 
\citep{Duffell2020}. Interestingly, the latter work finds that between $0.01<q<0.03$ the torque follows the trend we see here: fast accretion rates result in a stronger, more negative torque.

\subsection{Torque evolution}

Does the torque on an embedded IMRI change in response to the increasingly rapid inspiral?   
This is important as a physics question about the system, as well as a practical question about interpreting any future LISA data.  Additionally, it is important to know whether the numerous previous studies of disc torques for non-migrating satellites hold for GW-driven E/IMRIs, or if the torques are modified and need to be re-computed in simulations that include the GW-driven inspiral. 

The short answer is: it depends. Whether the GW-driven inspiral produces a significant effect on the 
torque depends on the mass of the inspiraling BH, the disc viscosity, Mach number, and the BH accretion efficiency. 

During the inspiral, we observe three effects that are possibly correlated with the nonzero inspiral rates. 
First, for all simulated binaries, the torque on a migrating BH shows long-term oscillations 
(on a timescale of $\sim100$s of orbits). 
 These are most clearly seen in Fig.~\ref{fig:Ttotal_adot} for $q=3\times10^{-4}$ and $q=10^{-4}$.  
Despite differing mass ratios these oscillations occur on similar timescales, which suggests they are caused by the same mechanism, likely global perturbations induced by the moving perturber. 
Although long-term modulations have been seen in the accretion rates of non-inspiraling, near-equal-mass binaries (e.g. see Fig.11 in \citealt{Dorazio+2013}), these modulations have not been seen for $q\ll 1$.
In our case, the long-term modulations in the torque arise from global perturbations as they are present in the torque 
contribution from outside the Hill radius. 
The magnitudes are of order $\lesssim10\%$, so for detectability purposes the torque can 
be approximated by an average, or possibly a stationary approximation for Type~II torques.
However, analytical approximations neglect the ability for Type~II torques to become positive.

Second, depending on the mass ratio, a sufficiently fast inspiral rate can change 
the magnitude of the torque. These changes arise from changes in the gas distribution inside 
the Hill radius. 
For our fiducial viscosity $\alpha=0.03$ (Fig.~\ref{fig:Ttotal_adot}),
the torque begins to deviate from its steady-state value\footnote{We note that the 
decreasing trend in the gas torque reported in Paper I was in fact a long-term transient 
due to the simulation having a closer outer boundary ($r/r_0 \le 2.75$, rather than 6).
Our slower inspiral runs confirm that 
the normalized torque on a $q=10^{-3}$ IMRI is indeed constant (albeit with oscillations) 
until reaching separations near $\sim 4 r_{\rm S}$, although this depends on the accretion rate.} 
at inspiral rates approaching $\dot{r}/{\Omega r \sim 10^{-4}}$. 
In reality, this rate is only reached in a circular binary when $q\ge10^{-3}$. Lower mass 
ratio binaries merge before reaching this rate. However, this trend is dependent on accretion
efficiency and disc viscosity. In discs with lower viscosity (or less efficient accretion onto the secondary),
we expect this deviation to occur at earlier times. Overall we find that the most prominent torque evolution
 occurs for a relatively massive satellite with $q=10^{-3}$, due to the significant pile-up of gas in its Hill sphere.
 Conversely, for mass ratios below $q\le3\times10^{-4}$, which do not show dominant Hill sphere torques and do not reach as fast inspiral rates, 
 one can safely estimate the average torque with analytical predictions for
 perturbers on fixed (circular) orbits
 and neglect any effects from the GW inspiral. We caution, however, that this only applies to the parameter space explored in this work, and disks with higher Mach numbers may lead to stronger deviations from analytical predictions. 
 
Interestingly enough, our $q=3\times10^{-4}$ simulation in the  $\alpha=0.03$ disc experiences a \emph{sign change} in 
the torque as it approaches merger. In this case, gas torques would initially slow down the 
inspiral before accelerating it,   
and in principle the net effect of gas torques in the GW signal (discussed below in Section~\ref{sec:LISA}) may be diminished. 
We do not expect such an occurrence to be common, given that 
for the majority of our simulations, torques do not--on average--change sign. Furthermore the sign change occurs at migration rates near the ISCO (see Fig.~\ref{fig:Ttotal_adot}), where as the EMRI spends most of its time (and thus is more likely to be observed) at larger separations where the torque remains positive. 
 For reference, we note in the last column of Table~\ref{table:parameters} which simulations show significant deviations from steady-state torques prior to merger.

Finally, the inspiral has an impact on torque fluctuations. In particular for $q=10^{-3}$, for which
gas pile-up on the BH is most significant, the Hill sphere torque shows an increase in 
fluctuation
amplitude
in the final $\sim1000$ orbits. With a higher Mach number, these fluctuations are more extreme and arise at earlier times.
In particular
the spikes in the Hill torque for $q=10^{-3}$ and $\mach=20$ may signify an 
interesting dynamical interaction  
between the BH's orbit and the gas, given that they occur at the same radii 
regardless of viscosity or the BH sink rate.
In simulations with varying boundaries ($0.4-0.5 \le r/r_0\le 3.0-6.0$) 
or higher resolution (800, rather than 666 radial cells), we find 
that these spikes in the torque
still occur at the same physical radii.

\begin{table}
\begin{tabular}{llc}
\vspace{-0.4cm}
\\
\multicolumn{3}{c}{\sc Fiducial Parameters and LISA Specifications}\\
\specialrule{0.8pt}{1pt}{1pt}
\vspace{-0.4cm}
\\
$z$  & Redshift  & $1$   \\
$M$  & Primary Mass       & $10^6 M_{\odot}$    \\
$\rm \tau$ & LISA mission lifetime & $4$ yrs \\ 
$\rm L$ & LISA arm length & 2.5 million km \\
$\rm N$ & Number of laser links & 6 \\ 
\specialrule{0.8pt}{1pt}{1pt}
\end{tabular}
\caption{LISA parameters are used when computing detectability.}
\label{table:LISAparameters}
\end{table}

\section{Significance for LISA inspirals}
\label{sec:LISA}

For the rest of the paper we take the results from our simulations, 
primarily the average torques measured in the $\alpha=0.03$ and $\alpha=0.1$ runs, and 
analyse their detectability in the GW signal.
First we introduce some relevant quantities that describe a GW event in the LISA band, with the goal of computing the detectability of the imprint of the gas torques.

As illustrated in Figure~\ref{fig:adot_a_q}, the inspiral rate for each mass ratio 
corresponds to a physical separation and gravitational wave frequency. 
Recall that the GW frequency is twice the orbital frequency for a binary on a circular orbit:
$ f = 1/\pi \sqrt{GM/r^3} $, 
and we have chosen these inspiral rates to correspond to IMRIs in the LISA frequency band. 

 The GW amplitude of a circular inspiral depends on the source distance (or redshift $z$), its frequency,
and its chirp mass $\mach_{\rm c}=M_1^{3/5} M_2^{3/5} / (M_1+M_2)^{1/5}$. 
The sky- and polarization-averaged GW strain amplitude of a source at co-moving coordinate distance $r(z)$ is given by 
\be\label{eq:h}
h = \frac{8 \pi^{2/3}}{10^{1/2}} \frac{G^{5/3} \mach_c^{5/3}} {c^4 r(z)} f_r^{2/3},
\ee
where $f$ is the observed GW frequency and $f_r \equiv f(1+z)$ is the GW frequency in the source's rest frame.

\begin{figure}
\begin{center}
\includegraphics[width=.5\textwidth]{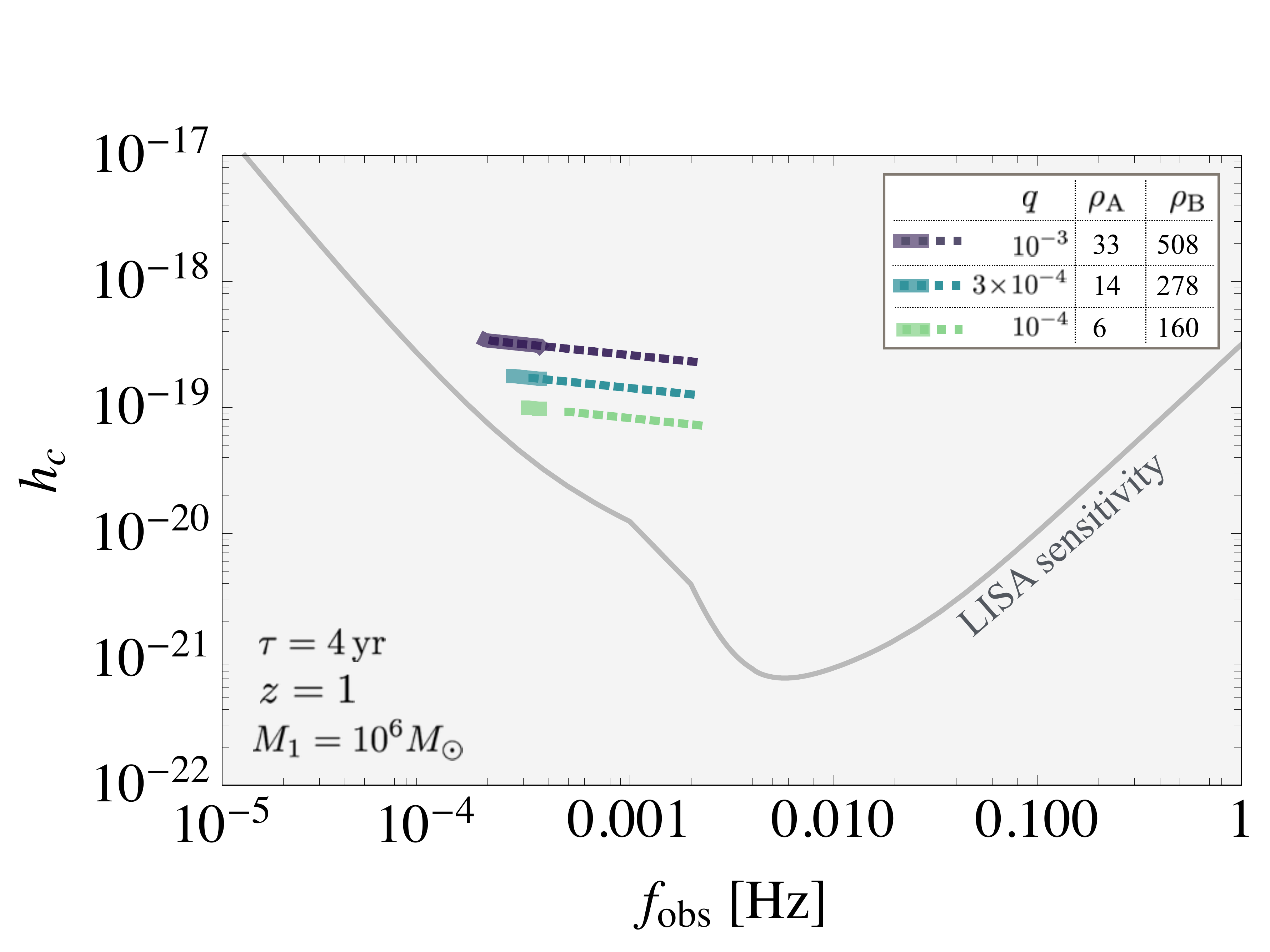}
\caption{The characteristic strain as a function of observed frequency against the dimensionless LISA sensitivity. 
For each mass ratio, we show two possible observations of duration $\tau=4$ years, using our fiducial parameters 
$M_1 = 10^6 M_{\odot}$ and $z=1$. 
Solid lines (observation `A') correspond to an earlier evolutionary stage of a binary that reaches a rest-frame separation of $10 r_{\rm S}$. Dashed lines (observation `B') cover the final coalescence of the binary, that reaches a rest-frame separation $r_{\rm ISCO}$. In the legend, we note the total SNR accumulated in each of these observations at early stages ($\rho_{\rm A}$) and later stages ($\rho_{\rm B}$), computed from Eq.~\ref{eq:totalSNR}. 
 The noise function is provided by \citealt{Klein2016}.}
\label{fig:strains}
\end{center}
\end{figure}

The \emph{characteristic} strain $h_c$ of a source whose frequency evolves 
during a LISA observation of time $\tau$ (i.e. the LISA lifetime) is given by 
$h_c = h\sqrt{n}$, where $n \equiv f^2/\dot{f}$, 
 which is a measure of the total number of cycles the source spends at each frequency
 (see \citealt{Sesana2005} for a more detailed
 discussion).
 In Fig.~\ref{fig:strains} we
 plot the characteristic strain of IMRIs at each simulated mass ratio for two different 
 possible observations of duration $\tau=4$ years, 
 the currently planned nominal LISA mission lifetime \citep{AmaroSeoane2018}. We assume a fiducial primary
 mass $M_1 = 10^6 M_{\odot}$ and place the source at redshift $z=1$.  These parameters 
 are listed in Table ~\ref{table:LISAparameters}.
The dashed lines correspond to the final $4$ years of a binary prior to reaching ISCO at $r_{\rm ISCO}=3 r_{\rm S}$,
and the shorter, solid lines correspond to binaries observed for 4 years and ending up at $10 r_{\rm S}$ (in their rest frame).

The signal to noise ratio (SNR) of the event is a measure of its "loudness", 
or a way to characterise its detectability compared to the LISA noise. 
It is an integral of the strain amplitude over the noise, given in the stationary phase approximation by
\be
\label{eq:totalSNR}
\rho^2 = 2 \times 4 \int_{f_{\rm min}}^{f_{\rm max}} df \frac{|\tilde{h} (f)|^2}{S_n^2(f) f^2}.
\ee
The pre-factor of two 
assumes the currently planned configuration of six links (effectively two independent interferometers), 
and $S_n(f)$ is the 
spectral density of LISA's noise 
per frequency bin, adopted from 
\citet{Klein2016}, where we assume 
an arm length of $2.5$ million km 
and include an estimate of confusion noise produced by foreground sources (galactic binaries).

For a fixed primary mass, binaries with lower mass ratios (smaller $\mach_{\rm c}$) emit 
weaker gravitational waves, and thus span shorter frequency windows during a fixed observation
time $\tau$. This reduces the total SNR as well as the chances of detecting a deviation in the signal.

\subsection{The imprints of gas on GWs}

\subsubsection{Phase drift}

Depending on the mass ratio and viscosity, gas torques either slow down or speed up the inspiral. 
For a gravitational wave event in the LISA data stream, this produces a phase drift in the waveform compared to that in vacuum, and a shift in the total 
phase accumulated during the event. 
If the accumulated phase (often defined in Fourier space) due to GW 
emission alone is $\Phi_{\rm GW}(f)$, the phase of an event experiencing gas torques will be 
$\Phi_{\rm GW}(f) + \delta \phi(f)$, i.e. the underlying vacuum signal plus some small deviation 
which can also depend on frequency. 
If the phase deviation $\delta \phi(f)$ is significant (and unique), the gas imprint is 
potentially distinguishable from the vacuum waveform (and from other deviating effects). This depends on
(i) the strength of the gas torque compared to GWs, which changes with radius; 
(ii) the frequency window that is observed, 
and (iii) the signal to noise ratio (SNR) of the event itself.  
Thus $\delta \phi$ is not only a function of frequency, since torques evolve with radius, but also scales with the disc surface density as this determines the strength of the torque.

For calculating the phase shift induced by gas we take the same approach as in Paper I 
(\citealt{Derdzinski2019}), but implement the updated result for the gas 
torque and apply this to all simulated mass ratios.
Recall that the total accumulated phase (in radians) of an event over a rest frame separation window is 
\be
\Phi_{\rm tot} = -2 \pi \int_{r_{\rm f}}^{r_{\rm i}} \frac{f}{\dot{r}} dr,
\ee 
where we assume the inspiral remains circular, and $\dot{r}$ is the inward radial acceleration. 
For an event in vacuum, this is purely driven by GWs and $\dot{r} = \dot{r}_{\rm GW}$.
In a disk, gas torques also affect the orbital evolution and $\dot{r} = \dot{r}_{\rm GW} + \dot{r}_{\rm gas}$. 
Note that because our measured torques on average do not vary significantly throughout the inspiral, we assume that these components are not coupled. 
The phase shift of a gas-embedded event is then given by  difference between the phase integral in vacuum and the observed accumulated phase:  $\delta \phi = \left| \Phi_{\rm tot}(\dot{r}_{\rm GW}) - \Phi_{\rm tot}(\dot{r}_{\rm GW} + \dot{r}_{\rm gas}) \right|$.
Given that gas torques are much weaker than GWs  ($\dot{r}_{\rm gas} \ll \dot{r}_{\rm GW}$, see Fig.~\ref{fig:Tgw}), we can simplify the latter integral and solve for the phase shift 
(or phase drift, as it accumulates) by
\be
\label{eq:delphi}
\delta \phi = 2 \pi
\int_{r_{\rm i}}^{r_{\rm f}}\! \frac{f(r) \, \dot{r}_{\rm gas}} {\dot{r}^2_{\rm GW}} \left [1+\mathcal{O}\left( \frac{\dot{r}_{\rm gas}}{\dot{r}_{\rm GW}}  \right)^2 \right] \, dr,
\ee
where $\dot{r}_{\rm gas} = 2 T_{
\rm gas} M_2^{-1} r^{1/2} (G M)^{-1/2}$ is the change in separation due to gas torques. 
Note that $\dot{r}_{\rm gas}$ can be a function of radius, 
as we define explicitly below, but that 
this function can vary for systems with different mass ratios or disc parameters. 

As shown in our fiducial runs in Fig.~\ref{fig:Ttotal_adot}, the magnitude of the gas 
torque when normalized by $T_0$ is approximately constant throughout the inspiral for all
mass ratios.  To compute the phase shift we use the \emph{average} of these torques, calculated over 
the last 2000 orbits of the slower-inspiral simulations. 
We neglect the 
short time-scale
oscillations in the torques, as well as the deviations from the average 
that occur at the highest
inspiral rates. Given that normalised torques either stay constant or 
increase in absolute
value with inspiral rate, this provides a lower limit on detectability. 

If $T_{\rm g}/T_0$ is constant throughout the inspiral, then torques scale with the radial dependence of $T_0$, which is dependent on the initial disc density profile. We can define the average gas torque analytically as 
\be
\label{eq:Tfit}
\langle T_{\rm gas}\rangle= C_{\rm fit} T_0(q,r,\mach,\alpha, \Sigma(r)),
\ee
where $C_{\rm fit}$ is the (constant) average of the torque before it begins to deviate due
to rapid inspiral. These fits are shown as horizontal dashed lines in Fig.~\ref{fig:Ttotal_adot} 
and also provided in Table~\ref{table:parameters}.

The gas torque on the satellite BH can be expressed in terms of the rate of change of specific angular momentum 
$\dot{\ell}_{\rm gas} = T_{\rm gas}/M_2$, which relates to the rate of change of separation
as $\dot{r}_{\rm gas}(r) = \frac{2}{\sqrt{G M_1}} r^{1/2} \dot{\ell}_{\rm gas}(r)$. Plugging 
this expression into Eq.~\ref{eq:delphi} allows us to solve for the shift in GW phase due to 
the gas torque on each binary, provided an observed frequency window (and corresponding range of separations). 
Note that in converting the torque from code units to physical units, these quantities must be scaled with our fiducial parameters by 
$T_{\rm gas} = T_{\rm code} \times G M_1 \Sigma_0 r_0 $
and  $\dot{\ell}_{\rm gas} = T_{\rm code} \times G \Sigma_0 r_0/q$.

We integrate over the two different frequency windows 
for each mass ratio, defined by the $4$-yr
observations shown in Fig.~\ref{fig:strains}. 
Since our simulated inspiral does not cover 
the entire observed frequency range, we extrapolate 
the torque fit from Eq.~\ref{eq:Tfit} to
the earlier stages, which implicitly (and reasonably) assumes that torques continue to scale with $T_0$ at earlier times.

\begin{figure}
\begin{center}
\includegraphics[width=.5\textwidth]{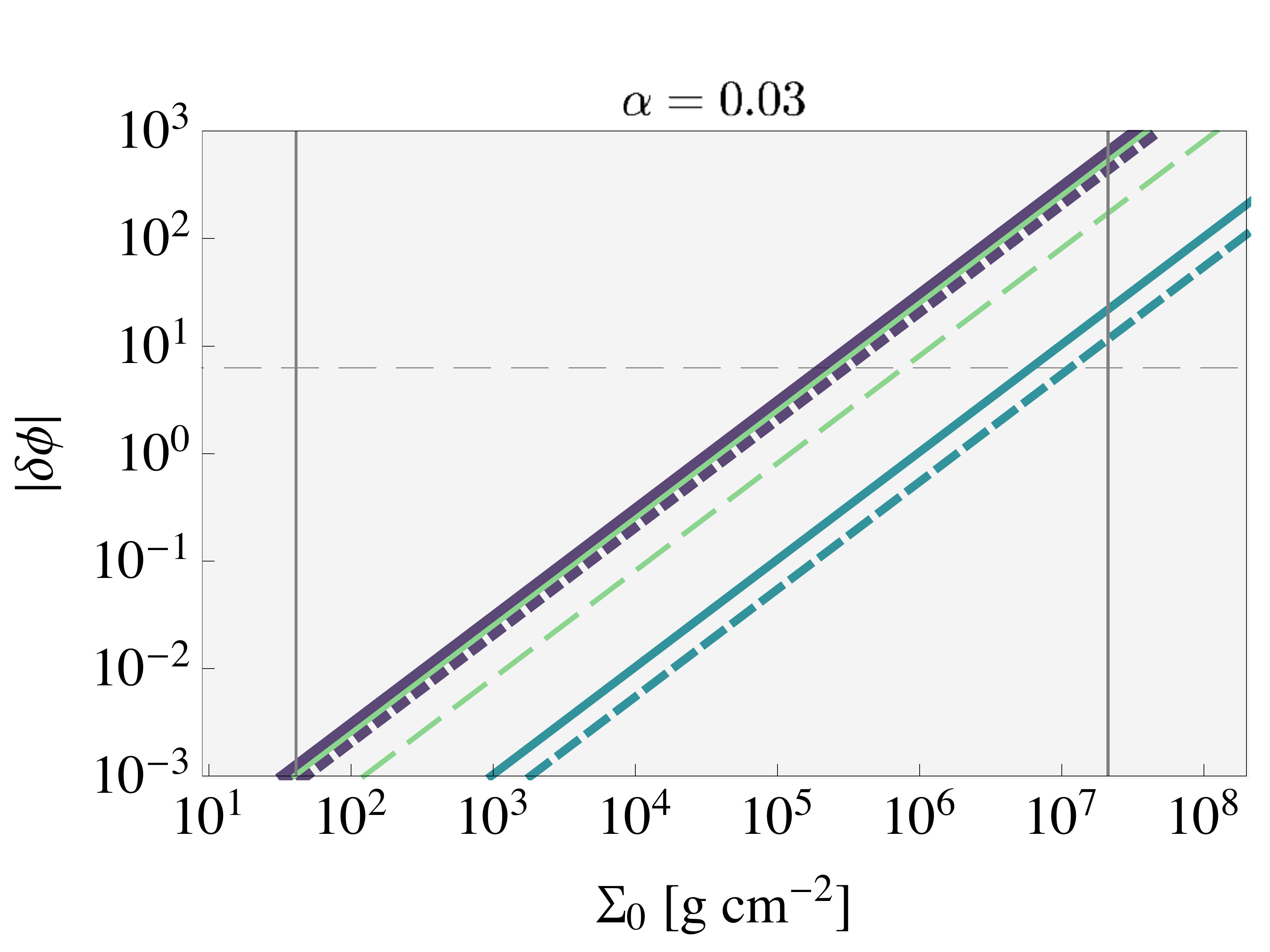}
\caption{Absolute value of the total accumulated phase shift (Eq.~\ref{eq:delphi}) for each observation window depicted in Fig.~\ref{fig:strains}, due to the average gas torque in the fiducial $\alpha=0.03$ runs. 
The colors and line types correspond to the same set of runs as in Figure~\ref{fig:strains}, covering a pair of observational windows for three different mass ratios.
Note that $\delta \phi$ can be negative or positive, depending on the sign of the torque. 
Vertical lines mark estimates for the disc surface density in the $\alpha-$disc and $\beta-$disc models at $r_0=3 r_{\rm S}$.
The dashed horizontal line delineates when $\delta \phi = 2\pi$, above which the shift exceeds an entire period. }
\label{fig:phaseshift}
\end{center}
\end{figure}

In Fig.~\ref{fig:phaseshift}, we plot the total accumulated phase shift for each observation as a function of the disc surface density. $\delta \phi$ scales linearly with $\Sigma_0$,
but accumulates to different values depending on the strength of the torque and the frequency window observed (both of which depend on $q$). For reference, we mark the surface densities in the $\alpha$- and $\beta$-discs at $r_0=3r_{\rm S}$ by
vertical
lines. In the low-density limit, gas torques only impart a phase shift of $\delta \phi \lesssim 10^{-3}$, which is likely undetectable. For higher densities ($\Sigma_0\gtrsim 10^5~{\rm g~cm^{-2}}$), the phase shift can exceed several radians.

\subsubsection{Detectability of waveform deviation}

The detectability of a deviation
can be estimated by calculating the SNR of the difference between the two waveforms ($\delta \phi$).
Similarly to Eq.~\ref{eq:totalSNR}, we calculate the SNR of the \emph{deviation} as
\be
\label{eq:SNRdev}
\rho_{\delta \phi}^2 = 2 \times 4 \int_{f_{\rm min}}^{f_{\rm max}} df \frac{|\delta \tilde{h} (f)|^2}{S_n^2(f) f^2},
\ee
where instead of the strain amplitude we integrate the strain deviation in Fourier space,
\be
\label{eq:deltah}
|\delta \tilde{h}|^2 = |\tilde{h}|^2 \left(1 - e^{i \delta \phi} \right)^2
= 2|\tilde{h}|^2 \left(1 - \cos{(\delta\phi)} \right)
\ee
This assumes the chirp is slow and that gas only imparts a difference in GW phase and not amplitude, also known as the stationary phase approximation (see \citealt{Yunes2011} and \citealt{Kocsis2011}).

\begin{figure}
\begin{center}
\includegraphics[width=.5\textwidth]{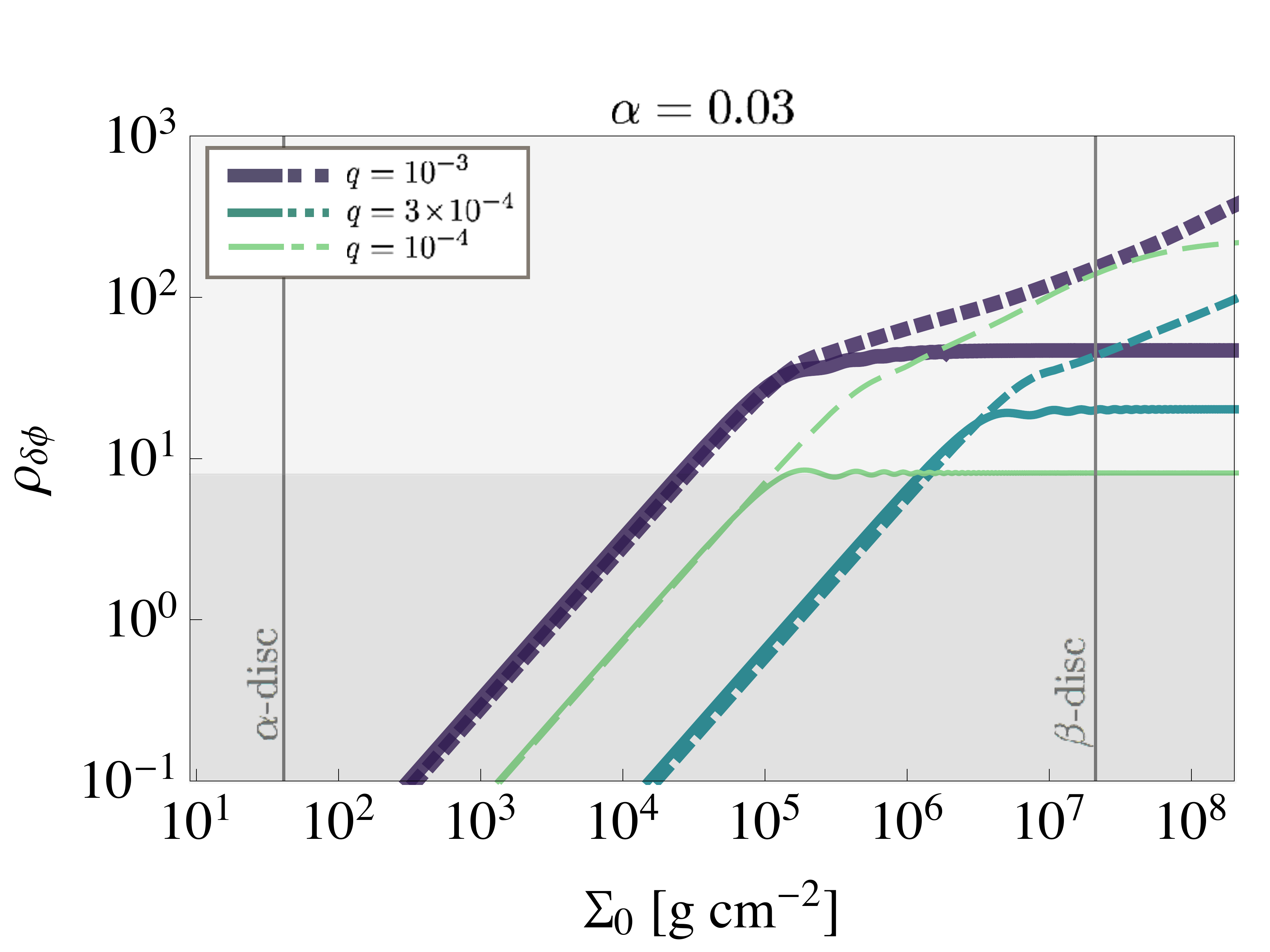}
\caption{Accumulated SNR of the waveform \emph{deviation} due to the average gas torque in our $\alpha=0.03$ runs, 
assuming 
binary parameters $M_1 = 10^6 M_{\odot}$ and $z=1$ and $\tau = 4$ years. Observational windows correspond to those depicted in Fig.~\ref{fig:strains}.
Observations of binaries
at lower frequencies (solid lines) saturate once the phase shift reaches $\delta\phi\approx2\pi$.
Binaries at later stages of coalescence continue to accumulate 
a larger phase shift due to the 
changing frequency, allowing higher values of disc density to yield large deviations. 
The shaded region corresponds to $\rho_{\delta \phi} < 8$. The phase shift is most detectable for binaries that are chirping significantly in dense discs, although the exact threshold varies for each $q$.}
\label{fig:SNR_allqs}
\end{center}
\end{figure}

We show the accumulated SNR of the gas-induced deviation for the $\alpha=0.03$ runs in 
Fig.~\ref{fig:SNR_allqs} as a function of disc surface density normalisation $\Sigma_0$.  
This allows us to assess, given a disc density, how distinguishable the phase-shifted waveform 
will be from the vacuum waveform. 
Just as the phase shift depends linearly on the surface density, $\rho_{\delta \phi}$ initially 
scales linearly with $\Sigma_0$. 
This can be interpreted from Eq.~\ref{eq:deltah}, where for small $\delta \phi$ the strain deviation becomes linear with the phase shift, as $|\delta \tilde{h}| \propto (1-\cos{\delta \phi})^{1/2} \approx \delta \phi$. 
However, at high enough surface densities, where torques shift
the phase by a whole period ($\delta \phi \gtrsim 2\pi$), this linear dependence disappears and the SNR saturates. 
This behavior is observed in binaries that are essentially monochromatic in frequency, an inevitable
feature of circular, extreme mass ratio inspirals that coalesce very slowly. 
The exception is for binaries that are approaching merger, 
whose phase shift accumulates 
past $2\pi$ as they sweep through higher frequencies. 
If an IMRI is embedded in a disc
with surface densities reaching that of the $\beta$-disc model,
 and we observe 
the late stages of coalescence, its waveform may be significantly altered.

\begin{figure}
\begin{center}
\includegraphics[width=.5\textwidth]{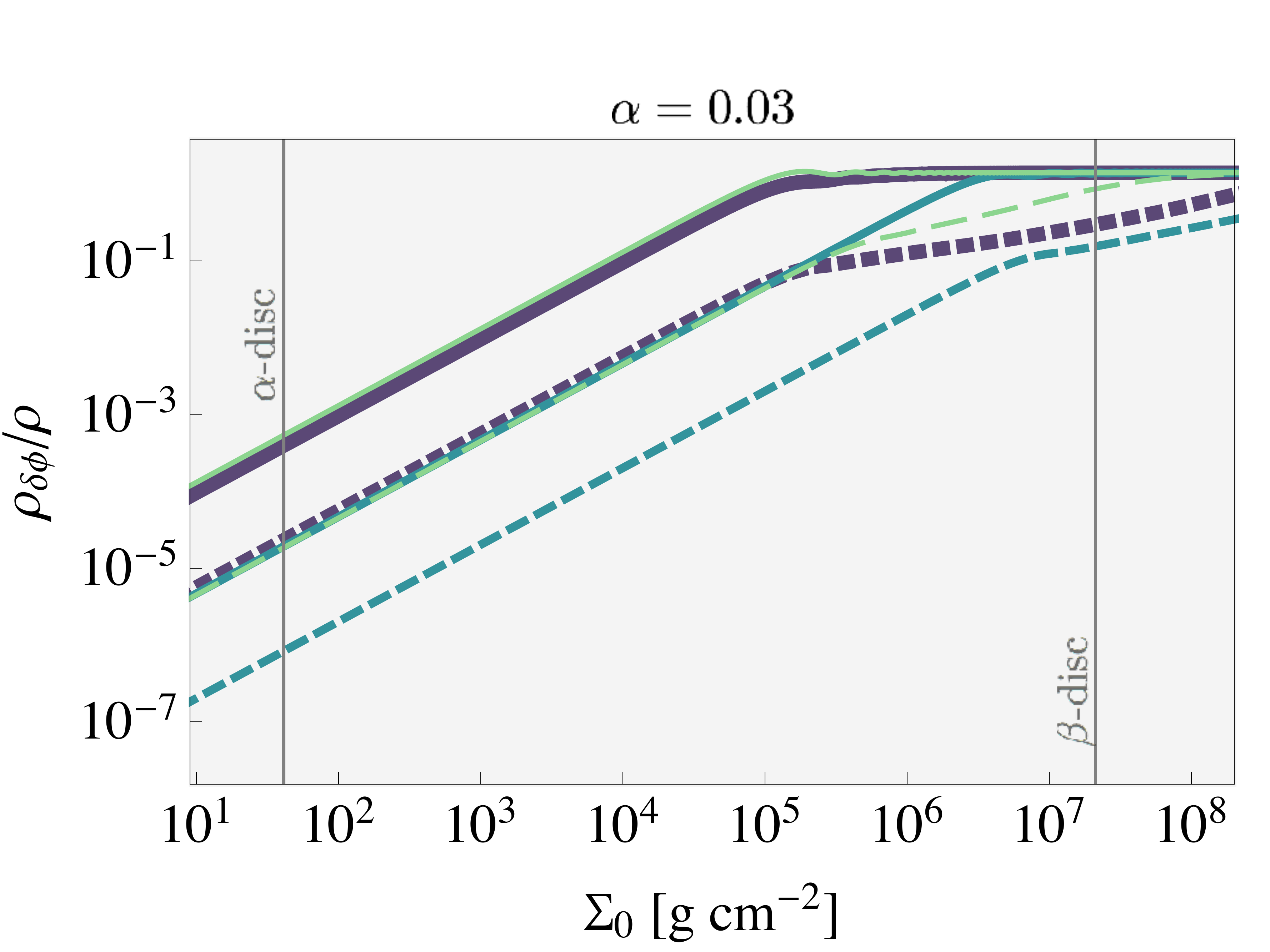}
\caption{Relative SNR (i.e. the SNR of the deviation divided by the total SNR of the event) for the $\alpha=0.03$ runs, with the same color key as in Fig.~\ref{fig:strains}. 
This shows that a faster chirp signal is relatively less affected by the gas torques, but the gas imprint is still more detectable because of the higher total SNR of the event (seen in Fig.~\ref{fig:strains}).
}
\label{fig:SNRrel}
\end{center}
\end{figure}

Ultimately the detectability of a deviation relies on the event accumulating substantial total SNR. 
Indeed, the observations for which a phase drift accumulates the highest $\rho_{\delta \phi}$ are those that accumulate highest total SNR, as indicated in Fig.~\ref{fig:strains}.
In Fig.~\ref{fig:SNRrel}, we show the SNR of the deviation divided by the total SNR of the event (a relative SNR, $\rho_{\delta \phi} / \rho$), as a function of disc surface density. 
This illustrates that 
the accumulated deviation for inspirals that are chirping faster (dashed lines) is weaker than for the inspirals that are observed at earlier times (solid lines). 
Nevertheless, we see in  Fig.~\ref{fig:SNR_allqs} these weaker deviations are more detectable due to the larger overall SNR.
Figure~\ref{fig:strains} clearly illustrates the reason: these tighter binaries are observed at frequencies close to the minimum of LISA's sensitivity curve.
Louder events $-$ ideally intermediate mass ratios at low redshift $-$ are the most promising candidates for detecting gas imprints.

Adopting a detectability threshold of $\rho_{\delta \phi} \gtrsim 8$, 
we conclude that 
the gas imprint is detectable for all simulated mass ratios 
if the disc density 
exceeds $\Sigma_0\gtrsim 10^{4-6} \, \rm g \, cm^{-2}$. 
 For an inspiral beginning at $10 r_S$, this threshold corresponds to a local disc mass of 
$M_{\rm enc} \sim \pi r^2 \Sigma_0 = 
 10^{-3}
  \left( \frac{r}{10 r_S(M_1)}  \right)^2
 \left( \frac{\Sigma_0}{10^5 {\rm \, g \, cm^{-2}}} \right)
 M_{\odot}$  within the vicinity of the secondary BH. 
The surface density required for detectability 
depends on the strength of the torque, which 
varies for each value of the mass ratio.\footnote{Compared to Paper I, the phase shift for $q=10^{-3}$ requires 
a slightly higher $\Sigma_0$ for detectability. This is primarily because  we fit $T_{\rm g}/T_0$ to a constant (rather than the decreasing polynomial shown in Paper I) which results in a weaker estimate for the torque strength. Additionally, we are using a more conservative observation time of 4 years (rather than 5).} 
Given that the gas torque on the $q=3\times10^{-4}$ binary 
is an order of magnitude weaker than for
the higher mass ratios, it requires a correspondingly 
higher $\Sigma_0$ for detectability. 
The detectability of gas for lower mass ratios at 
earlier stages suffers from weaker GW emission 
and more modest frequency evolution, and gas torques are less detectable even for the highest disc densities.

\begin{figure}
\begin{center}
\includegraphics[width=.5\textwidth]{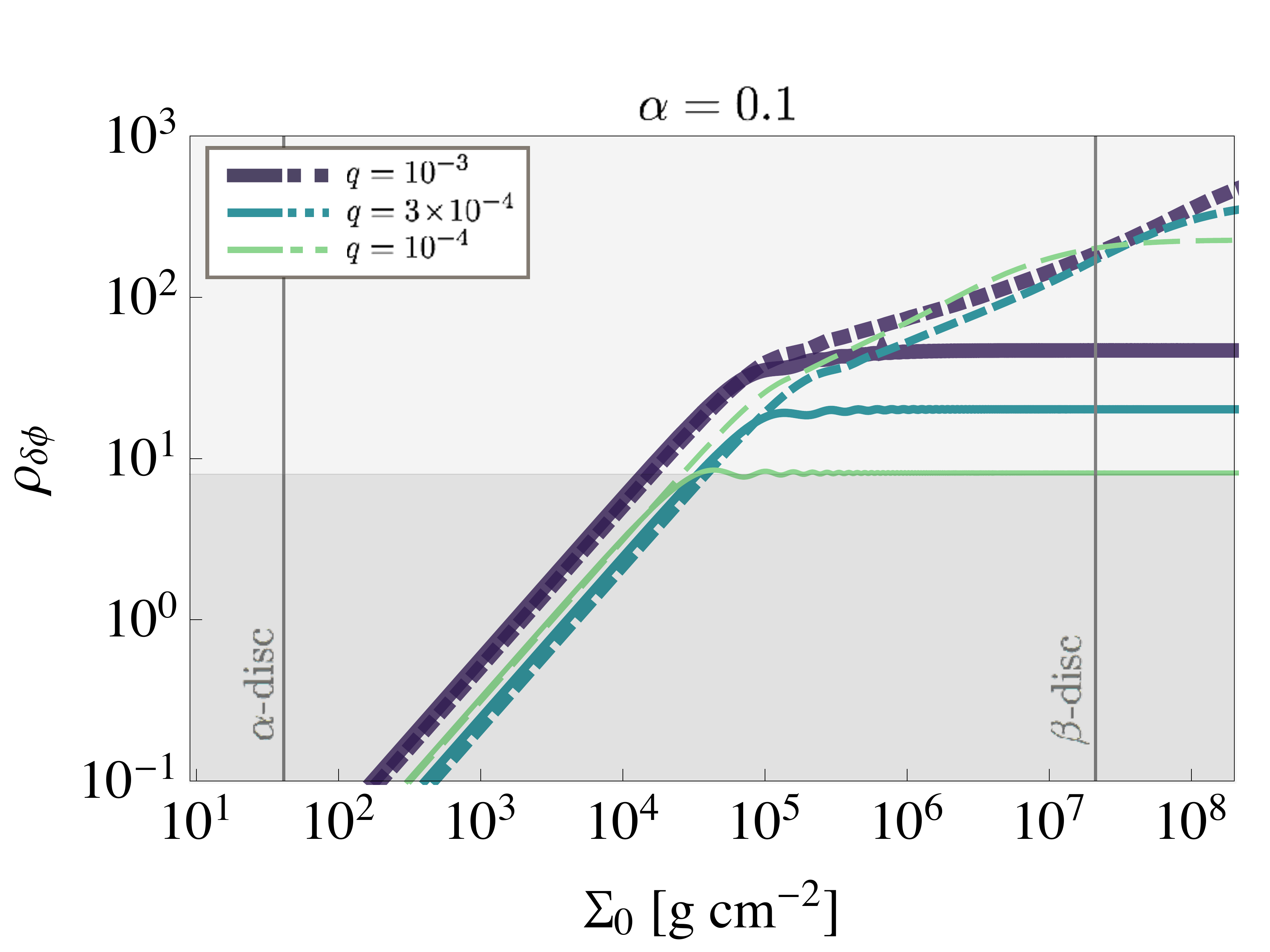}
\caption{Accumulated SNR of the deviation due to average torques in the $\alpha=0.1$ runs, shaded up to our detectability threshold of $\rho_{\delta \phi} \ge 8$.
Higher viscosity 
generally produces stronger torques; hence the imprint 
is detectable at lower disc densities.}
\label{fig:SNR_allqsa0p1}
\end{center}
\end{figure}

For higher viscosities, where torques are stronger for each mass ratio, 
the detectability of a 
deviation is improved (see Fig.~\ref{fig:SNR_allqsa0p1}), 
and an SNR deviation of $\rho_{\delta \phi} \gtrsim 8$ can be reached 
with lower disc densities, $\Sigma_0\sim 10^{4} \, \rm g \, cm^{-2}$.

\begin{figure}
\begin{center}
\includegraphics[width=.5\textwidth]{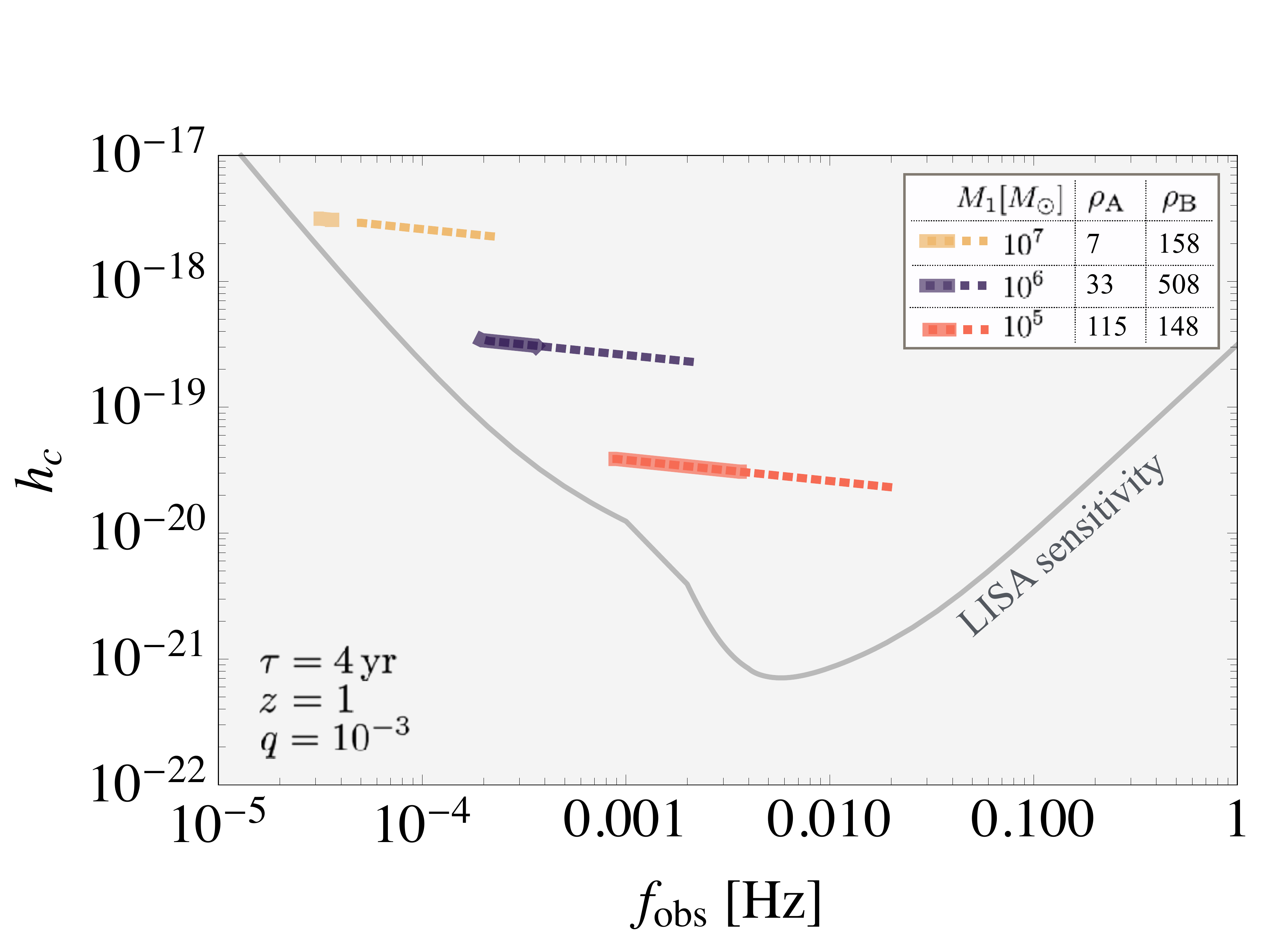}\\
\includegraphics[width=.5\textwidth]{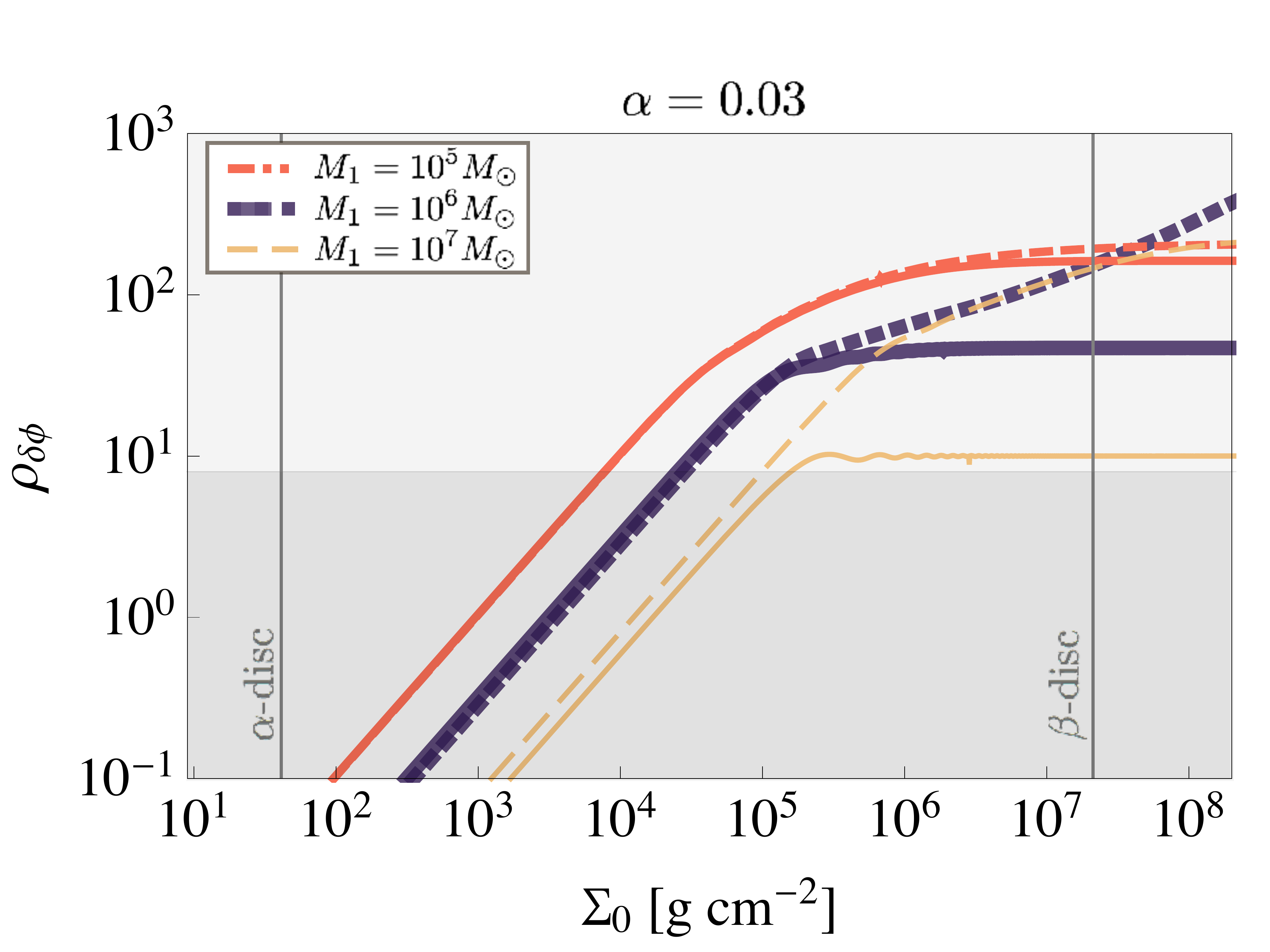}
\caption{\emph{Top panel}: Characteristic strain {\it vs.} observed frequency for $4$-year observations of 
$q=10^{-3}$ binaries at $z=1$, varying the 
primary mass $M_1$ 
from $10^5 M_{\odot}$ to $10^7 M_{\odot}$.
Solid lines correspond to early stages of evolution where the binary reaches a rest-frame separation of $10 r_{\rm S}$. Dashed lines correspond to observing the final coalescence, ending when the binary merges at $r_{\rm ISCO}=3r_{\rm S}$. In the legend we provide the accumulated SNR for the early (`A') and late (`B') observations, respectively. 
For lower binary mass, the merger occurs 
at a smaller $r_{\rm ISCO}$ and correspondingly higher frequencies. 
This affects the detectability 
of the event. 
\emph{Bottom panel}: SNR of gas deviation as a function 
of disc surface density given
the observation windows depicted above, 
and using the average 
dimensionless 
gas torque from the $q=10^{-3}$ and $\alpha=0.03$ runs. 
A binary with 
$M_1=10^5 M_{\odot}$ accumulates a stronger waveform deviation
as it spans a larger frequency range.
Note that the vertical line for the 
$\Sigma_{\beta}$ estimate corresponds to a disc around 
an $M_1=10^6 M_{\odot}$ BH; see Eq.~\ref{eq:Sigma_beta}
for the weak scaling of the disc surface densities with BH mass. }
\label{fig:SNRrel_masses}
\end{center}
\end{figure}

One may wonder how our choice of primary mass affects the detectability of the gas imprints. 
In principle, IMRIs can 
occur for more or less massive primary MBHs 
while still emitting GWs within the LISA frequency band.
We demonstrate the effect of our choice of 
primary mass $M_1$ for a fixed mass ratio $q=10^{-3}$ 
in Fig.~\ref{fig:SNRrel_masses}. 
Higher mass binaries emit louder gravitational waves, 
but they merge at lower frequencies due to their 
increasingly large ISCO ($r_{\rm S} \propto M$). 
Taking the mass ratio $q=10^{-3}$, we plot the strain and corresponding 
detectability of the phase shift,
adopting the dimensionless gas torque from our fiducial run 
(q1e3a03, where $\langle T_{\rm gas}\rangle/T_{0} = 0.21$). 
We fix each observation window to the binary 
reaching $3 r_{\rm S}$ in a $4$ year observation. In this 
case the lower-mass binary 
can exhibit more detectable gas torques, 
given that the coalescence occurs at frequencies
where LISA is most sensitive. 
However, the overall SNR of the event (and consequently the SNR of the deviation) will depend on the range of observed frequency of the binary and its relation to the peak sensitivity--notice that the coalescence of the $10^6 M_{\odot}$ IMRI attains the highest total SNR (dashed purple line, $\rho_{\rm B}$), 
while the $10^5 M_{\odot}$ merger occurs at higher frequencies, reducing the total SNR (dashed orange line, $\rho_{\rm B}$). 
Additionally, 
lower-mass binaries span a larger frequency 
range in a fixed observation time, simply because
at fixed $r/r_{\rm S}$ 
the frequency evolution rate $\dot{f}$ scales 
more steeply with frequency than chirp mass ($\dot{f}\propto M_c^{5/3}f^{11/3}$). 
In summary,
the detectability of the gas deviation is tied 
to the stage of the coalescence we observe -- binaries
that are chirping in the cusp of LISA's sensitivity are 
the most promising candidates for extracting a phase shift.

\subsection{Uniqueness and Degeneracies} 
It is important to consider whether the deviations we find in the waveform are degenerate with 
changes in the chirp mass or other system parameters, and/or with
other possible environmental effects, and if such degeneracies may hinder parameter estimation or chances of detection. 

A critical feature for distinguishing between various 
environmental effects and system 
parameters is the respective scaling of each effect 
with binary separation or GW frequency.
For example, some proposed modifications to general 
relativity predict waveform deviations 
that would increase as the EMRI coalesces, scaling with
a predictable power of frequency\footnote{However, one would expect effects from modified gravity to be present in \emph{all} E/IMRIs within a certain frequency range, while gas should only affect a subset of the events.} \citep{Yunes2009}. 
In principle a deviation to the waveform could be 
interpreted as a binary with different system
parameters that also determine the frequency evolution, 
such as chirp mass $M_c$. This can be distinguished
by how various parameters affect the waveform as the frequency evolves. 
This dependence is often 
quoted in Fourier space, where deviations can be 
compared to Post-Newtonian (PN) terms in the Fourier 
phase. The scaling of changes in parameters with 
frequency can then be compared amongst system 
parameters, external effects, or modifications of general relativity.

Neglecting short-timescale fluctuations, 
our simulated gas torque scales with the analytical estimate $T_0$, 
which is a function of the disc density profile. Given our disc model, 
$T_{\rm gas} \propto r^4 \Omega^2 \Sigma(r) \propto r^{1/2}$. 
The viscous torque also carries the same scaling with radius, given by 
$T_{\nu}\propto r^2 \Omega \nu(r) \Sigma(r) \propto r^{1/2}$. 
This predicts that gas torques get progressively weaker as the binary coalesces: if 
$f_{\rm GW}\propto r^{-3/2}$, then the gas torque initially scales with GW frequency as 
$T_{\rm gas} \propto f_{\rm GW}^{-1/3}$, and the integral for the deviation scales as 
$\delta \phi \propto f^{-13/3}$.
This frequency-scaling can be compared to the phasing function for a circular inspiral 
around a non-spinning BH, 
where 1.5 PN terms scale with frequency as $\phi\propto Af^{2/3} + Bf$ 
(where $A$ and $B$ are constants that depend on system parameters, 
see a more detailed description in \citealt{WillYunes2004}).
In this case, gas torques scale uniquely with frequency, 
suggesting that 
the impact of the gas torque
is not degenerate 
with system parameters or GR corrections. 
However, confirming a phase drift to be of gas-disc origin, rather than other possible environmental effects, 
may only be possible for events that span a sufficiently large range of frequencies.

These considerations exclude evolutions in the dimensionless torque that can occur at the fastest inspiral 
rates, particularly for low-viscosity discs, or if the BH migrates through a disc where 
parameters (e.g. $\alpha$, $\mach$) vary with radius. It also neglects the oscillations that occur in the torque which
we see for all simulated inspiral rates. If measurable, these time-variable fluctuations could
be a smoking-gun signature of disc response to an embedded IMRI. 
Ultimately the frequency dependence of the effect varies with disc physics, and changes in 
$\alpha$, $\mach$, or accretion efficiency will lead to different scalings.

\subsection{Caveats}
This study provides a crucial first step in determining how gas torques respond to a 
GW-driven inspiral, but much work must be done to make more accurate observational predictions. 
These simulations are numerically challenging in that 
they 
require a large boundary (to avoid transient effects) and high spatial resolution (to resolve the gas in the immediate vicinity of the satellite BH), they must model the global disc (the small-scale gas dynamics near the satellite depends on the global disc), and they must be evolved for several thousand dynamical times for each set of parameters (to avoid transient behaviour reflecting the initial conditions). 
In the interest of computational efficiency, we have neglected several physical processes that are important for modeling a realistic system. 
We summarize some of our limitations here.

Our simulations do not resolve 3-dimensional gas morphology which 
may be important for resolving accurate torques \citep{Tanigawa2012,Szulagyi2014,Morbidelli2014}, although in some parameter regimes 2D simulations are sufficiently accurate (e.g. see \citealt{Lega2015,Uribe2011}). 
In particular, the asymmetric density distribution that determines the gravitational torques is in some cases concentrated near or even inside the smoothing length.  The dense pile-up is only a factor of $\sim$two  inside the smoothing length in the most extreme case (i.e. for $q=10^{-3}$; see Fig.~\ref{fig:torq_grid}). It is useful to keep in mind that unlike in N-body or smoothed particle hydrodynamics simulations, where smoothing is done purely for numerical stability, the smoothing we employ here is physically motivated: the smoothing of the potential mimics a vertical averaging of the gravitational forces.  Nevertheless, the small-scale density distribution may be impacted by our choice of smoothing prescription, and 3-dimensional simulations will be necessary to fully understand how this morphology arises and if this gas distribution and the resulting torques are modified in 3D.

Our disc model is isothermal 
and does not include radiative cooling, heating, nor more sophisticated physics such as magnetohydrodynamics or radiation pressure, 
all of which may alter the gas dynamics near the BH.
Future work must consider how gas morphology near the BH is affected by accretion rate and feedback from the BH itself, 
which may heat the gas in its vicinity and dampen the torque \citep{Szulagyi2016}.
 While feedback tends to reduce the gas density in the vicinity of the migrator, it can do so assymetrically, producing an additional `heating torque' (as discussed in  \citealt{Hankla2020}).

The choice of accretion
prescription and sink timescale of the embedded BH should also be considered carefully. 
Our estimate assumes that accretion occurs on the viscous timescale via a thin disc around the BH.
However, given that the specific angular momentum of the gas with respect to the gap-opening
perturber is low (and the resulting accretion torque is negligible, as discussed in Paper I), 
perhaps a quasi-spherical accretion prescription (i.e., Bondi accretion; \citealt{Edgar2004})
 would be more appropriate. The possibility for super-Eddington accretion
  rates should be considered, which can result in feedback that further affects the orbital properties of the BH \citep{Gruzinov2020}. 
 We expect that the inclusion of such affects will affect the precise value of the torque and its variability, particularly for more massive IMRIs with $q\sim10^{-3}$ which experience the strongest nonlinear effects and more significant accretion/feedback. 
Nevertheless, we do not expect these effects to significantly change our detectability results of gas imprints, as our estimates utilize the time-averaged torques. 
Rather more sophisticated treatments may change predictions of the precise signature in the waveform - i.e. the sign, strength, and frequency-dependence of the phase drift, 
ad any effects that lead to a strong increase or reduction in torque strength (such as significant feedback), will alter the specific disc density constraints for detectability.

Finally, our simulations are purely Newtonian, neglecting any relativistic effects which can alter gas dynamics in the inner regions of the accretion disc closer to the primary BH. 
 However, we do not expect the inclusion of relativistic effects to substantially affect the detectability of the gas imprint, since the phase shift is mostly accumulated at larger separations. 
We assume the binary inspiral remains circular, when in fact gas discs may excite non-negligible 
eccentricity in the orbit (e.g. \citealt{Goldreich2003,DAngelo2006}),
and this eccentricity may also produce additional modulations in the torque. Additionally, our estimates of binary evolution may be slightly inaccurate due to our use of the \citet{Peters64} quadrupole formula for the inspiral rate, which is lacking PN terms that become important near the ISCO~\citep{Zwick2019}.

\section{Discussion}
\label{sec:discussion}

In the present work we analyse torque evolution during GW-driven inspirals in the 
intermediate mass ratio regime. These sources, while their rates are 
less certain, provide the tantalizing 
possibility of probing nonlinear binary+gas-disc physics. At lower mass ratios, 
they evolve more slowly and quietly than near equal-mass MBH mergers,
yet they maintain the ability 
to accumulate significant SNR.

 In all cases, we observe short time-scale modulations in the torque throughout the inspiral. 
For most cases, the strength of the torque is on the order of simple analytical predictions (some fraction of $T_0$), but the precise value and the sign of the torque changes nontrivially with $q$ and $\alpha$. 
For our highest simulated mass ratio $q=10^{-3}$, where gas pile-up on the BH is significant, torques 
are positive (outward), noisy, and show a distinct increase in variability at the fastest inspiral rates. 
The strength and evolution of the torque in this case is sensitive to accretion efficiency: less efficient accretion can lead to an increasingly positive torque as the binary coalesces.  
This effect may be amplified in more massive IMRIs, but this must be confirmed with future simulations. 
For lower-$q$ inspirals, 
we find that torques are smoother and become more \emph{negative} at fast inspiral rates, depending on the viscosity. 
Overall, IMRIs with different system
parameters can experience different torque evolution.

 Unsurprisingly, the dependence of torque with inspiral rate itself depends 
 on disc parameters, namely $\mach$ and $\alpha$. We find that inspirals in 
 low $\alpha$ discs show deviations 
 in the torque (compared to a constant dimensionless value) at earlier times. We interpret this as the 
 disc's inability to respond to the satellite BH's 
 increasing inspiral speed. Given the current
 understanding of viscosity in AGN discs (from 
 estimates of $\alpha$ in MHD simulations), such 
 low viscosities are unlikely. In the case of 
 higher $\alpha$, torques may hold a relatively steady 
 dependence on radius that scales with the viscous torque.

We find that fluctuations in the torque are highly dependent on the Mach number, or disc temperature. Hotter (low $\mach$) discs produce smoother, "well-behaved" torques, while thinner, colder discs 
exert torques that are stronger (scaling on average with the $T_0$ prediction) and more variable. Our simulations of a  $\mach=30$ disc yields a torque with strong variations that oscillate between positive and negative values. 
This suggests that IMRIs in thin, supersonic discs may experience the most dynamic gas effects.

The variability in torques throughout the inspiral may be attributed to interesting gas dynamics that warrants further investigation.
 For example, our $q=10^{-3}$ simulations show peaks in the noise amplitude at $r\approx8,6$ and $4 \, r_{\rm S}$ (most clearly seen in Fig.~\ref{fig:Thill}). These peaks occur at the same physical radii regardless of the simulation resolution,  boundary location, or sink prescription. 
 This raises the possibility that torques may show coherence
with binary separation. 
We defer this analysis, as well as a closer look at the underlying cause of torque variability, to future work.

Assuming a gas-embedded IMRI event has sufficiently high SNR and
spans a large enough frequency range that a frequency-dependent waveform deviation
$\delta \phi (f)$ can be measured,
it will not only confirm the deviation to be of gas origin
 (at the least ruling out other possibilities),
 it can also provide an invaluable 
 measure of disc properties as a function of radius.  
 While a measure of accumulated phase shift can place a 
 constraint on the disc density $\Sigma_0$, in the most
 optimistic case a phase \emph{drift} $\delta\phi(f)$
 could reveal how the surface density changes with
 radius $\Sigma(r)$, which is a distinguishing factor 
 amongst several accretion disc models. 
 Understanding the complex physical processes at play 
 in the inner regions of AGN accretion discs
 is an active field of research.
 Recent radiation, magneto-hydrodynamic simulations by \citet{Jiang2019} predict that inner 
 disc regions may have lower densities than predicted by the 
 $\beta-$disc model, 
 although the densities 
 should increase with accretion rate and may change with central MBH mass.

We note that these results are all subject to the limitations of our simplistic disc model, 
where the Mach number 
and aspect ratio do not vary with radius. 
This approach allows us to investigate whether changes 
in the torque are truly due to the GW inspiral, and not due to
encounters with varying disc dynamics. 
Previous works suggest that torques in response to an artificially imposed migration may change sign at fast migration rates \citep{Duffell2014}, but those simulations 
utilise a disc model with constant surface density, 
implying a radially dependent aspect ratio. 
Comparison between these types of studies highlights the 
importance of considering how different 
disc models may affect an inspiraling BH. 
On this basis, IMRIs embedded in more physically-motivated discs $-$
in which the Mach number and disc structure changes 
with radius $-$ may show more extreme changes in the torque,
although this remains to be confirmed with more sophisticated models.

The presence or absence of a phase drift should be 
considered in conjunction with other characteristics 
of the source, particularly any tell-tale signs of gas discs. 
A likely signature of a gas-embedded E/IMRI 
would be the combination of a phase drift with low eccentricity: gas-embedded E/IMRIs should be distinctly less eccentric than those expected to occur in dry galactic nuclei.
They may be close to circular, but with mild gas-driven eccentricity 
($e\lesssim0.2$, \citealt{Ragusa2018}, D'Orazio et al., in prep, Zrake et al. in prep) or eccentricity induced by other embedded perturbers. 
Additionally, if the accretion disc is aligned with the spin of the central MBH (this may only be true in some cases to varying degrees, see \citet{Volonteri2013} and references therein), the spins of the binary components may be closely aligned. However, this will depend on the history and nuances of accretion onto the disc-born BH, for which there are many uncertainties.

Our estimates of detectability of the deviation implicitly rely on the assumption that we have a waveform catalog
of all possible I/EMRIs. Currently, the catalog of intermediate mass ratio waveforms is 
incomplete \citep{MandelGair2009}. 
A lack of available waveforms will affect the accuracy with which we can extract binary parameters from the signal, let alone detecting subtle environmental deviations \citep{CutlerV2007}. 
To further complicate the picture, the range of possible AGN environments means that 
deviations may vary from system to system. In terms of data analysis, 
we suggest methods that search for \emph{generic} deviations in a waveform, which can be informed by studies such as that presented here.
In the optimistic case, one hopes that gas imprints can be traced back to constraints on the source environment. 
However, challenges will arise with circular, gas-embedded I/EMRIs that are in early stages of evolution, as these will arise in GW data as quasi-stationary sources. In these cases, assessing degeneracies, or disentangling the phase-shifted waveform between system parameters and various environmental effects, will not be possible if one cannot measure $\dot{f}$ of the source within a limited observation time.

Gas also provides the opportunity for electromagnetic emission 
that may coincide with the GW 
event. We do not address this issue here, but remind the reader that 
combining a phase shift with 
any associated EM signatures would be invaluable for confirming 
the presence of circumbinary gas and for learning 
about AGN discs. For example, the coalescence of a 
gap-opening secondary may accompany a change 
in AGN continuum that correlates with the mass ratio  
(e.g. \citealt{Gultekin2012}). 
 Variable signatures in the broad component of Fe K$\alpha$ line emission, believed to be associated with the inner disk structure, may also indicate the presence of a gap-opening IMRI via periodic oscillations \citep{2013MNRAS.432.1468M}.
Such signatures would be detectable with the upcoming X-ray observatory \emph{Athena} \citep{2020AN....341..224B}, and a powerful probe of inner disk morphology if observed concurrently with LISA \citep{2020NatAs...4...26M}.
These events may also be correlated with the fading observed in changing-look quasars, where the characteristic timescales suggest the change in emission is due to an abrupt change in the structure of the innermost accretion disc \citep{Stern2018}. Such changes could by driven by thermal or magnetic instabilities and possibly triggered by embedded perturbers.

\section{Conclusions}\label{sec:conclusions}

In this paper, we analyse the gas torques
on an intermediate mass ratio 
binary inspiral embedded in an accretion disc.
We present a suite of simulations of IMRIs embedded in two-dimensional, near-Keplerian,
isothermal accretion discs, where the satellite BH is modeled as a 
smoothed point-mass with a sink 
prescription. We analyse the torque exerted by the gas 
on the inspiraling BH for a range 
of mass ratios ($10^{-4}<q<10^{-3}$), disc viscosities ($0.01<\alpha<0.1$), and Mach numbers ($10<\mach<30$).
We also consider binaries at different states of the inspiral, and with and without allowing the satellite BH to accrete.
Here we summarize our conclusions:

\begin{itemize}
\item As in similar numerical studies, we find that torques 
in the intermediate mass ratio, 
gap-opening regime have a nonlinear scaling with disc properties. 
Torques either slow down or 
speed up the inspiral; their strength is some fraction ($\sim1 \% - 120\%$) 
of the Type~I torque $T_0$ \citep{Tanaka2002}, but 
these values are sensitive to small changes in $q$, $\alpha$, and $\mach$.

\item During the inspiral, the torques exerted by the gas on the satellite BH show weak fluctuations,
but the 
average strength of the torque ($T_g$ normalised by $T_0$) remains constant for the majority
of inspiral rates in the LISA band.
For the fastest inspiral rates, particularly for $q=10^{-3}$ approaching the ISCO,
torques exhibit an increase in variability originating
from the gas flow within the BH's Hill radius.

\item  We scale our simulation setup to a fiducial binary with primary mass 
$M=10^6 M_{\odot}$ at $z=1$ in order to compute the 
detectability of gas-induced deviations 
in the GW waveform. 
 Using the average of $T_{\rm gas}$ for each mass ratio, we
  compute the accumulated phase shift 
  and the corresponding SNR of the deviation as a function of disc 
  density normalisation. 
 We find that the phase shift is detectable (with relative SNR $\rho_{\delta \phi}>8$) when the source is embedded
in a disc with surface density $\Sigma_0\gtrsim10^{4-6} \, \rm g \, 
cm^{-2}$ (or a local disc mass $M_{\rm enc} \sim 10^{-3} M_{\odot} (\Sigma_0/10^5 {\rm g \, cm^{-2}})$)
,  
depending on the mass ratio and disc viscosity. 
Detectability is maximized for the loudest events that are chirping significantly throughout a LISA observation, 
and coalesce at frequencies near $\sim 5\times10^{-3}$ Hz where LISA is most sensitive.

\item This work is an important step towards 
understanding the scope of environmental impact on LISA sources due to circumbinary gas.
Ultimately the strength, direction, and evolution of the torque 
exerted on a gas-embedded IMRI is 
dependent on the mass ratio and disc parameters, 
and the resulting waveform deviations can manifest in a variety of ways. 
A measure of a phase drift can provide a constraint on the disc surface
density or disc structure, provided we have the tools to extract a variable 
deviation from the GW signal.  

\end{itemize}

\subsection*{Acknowledgements}
The authors thank the anonymous referee for insightful comments. 
AD acknowledges support by the National Science Foundation (NSF) Graduate Research Fellowship under Grant DGE 1644869.
DJD acknowledges support from the Institute for Theory
and Computation Fellowship.
ZH acknowledges support from NSF grant 1715661 and NASA grants
NNX17AL82G and 80NSSC19K0149.
Computational resources were provided by the NASA High-End Computing (HEC) Program through the NASA Advanced Supercomputing (NAS) Division at Ames Research Center.  We acknowledge the use of the following
\emph{software}:  DISCO \citep{Duffell2016}, Matplotlib \citep{Hunter2007}, NumPy \citep{numpy}, Mathematica 12.0 \citep{Mathematica}.

\subsection*{Data availability}
The simulation data that support the finding of this study are available from corresponding author AD, upon reasonable request.

\bibliographystyle{mnras}

\bibliography{paper}

\end{document}